\documentclass[aps,preprintnumbers,prd,twocolumn,superscriptaddress,nofootinbib]{revtex4-1}
\usepackage{blindtext}
\usepackage{lineno}
\usepackage{bm,color,xcolor}
\usepackage{slashed} 
\usepackage{graphicx}
\usepackage{soul} 
\usepackage{amsmath}
\usepackage{multirow}
\usepackage{float}
\usepackage{comment}
\usepackage{mathtools}
\usepackage{xspace}
\usepackage[colorlinks=true
,urlcolor=purple
,anchorcolor=purple
,citecolor=purple 
,filecolor=purple
,linkcolor=purple
,menucolor=purple
,linktocpage=true
,pdfproducer=medialab
,pdfa=true
]{hyperref}
\usepackage[nameinlink,capitalize]{cleveref} 
\usepackage{lipsum}

\makeatletter
\def\l@subsubsection#1#2{}
\makeatother

\makeatletter
\AddToHook{cmd/appendix/before}{\def\cref@section@alias{appendix}}
\makeatother

\newcommand{\GeV}{\ensuremath{\mathrm{GeV}}}
\newcommand{\TeV}{\ensuremath{\mathrm{TeV}}}

\newcommand{\beq}{\begin{equation}}
\newcommand{\eeq}{\end{equation}}
\newcommand{\bea}{\begin{eqnarray}}
\newcommand{\eea}{\end{eqnarray}}

\definecolor{blue-violet}{rgb}{0.33, 0.17, 0.89}
\newcommand{\gitlink}{\href{https://github.com/jchoi55/MINT}{{\large\color{blue-violet}\faGithub} \textsc{g}it\textsc{h}ub}\xspace}

\newcommand{\mint}{\texttt{MINT}\xspace}

\newcommand{\vect}[1]{\boldsymbol{#1}}

\usepackage{fontawesome5} 

\definecolor{orcidlogocol}{HTML}{A6CE39}

\newcommand{\myorcid}[1]{\href{https://orcid.org/#1}{\textcolor{orcidlogocol}{\faOrcid} #1}}

\begin{document}

\preprint{UMN-TH-4432/25}

\title{The Forward Neutrino Flux and its Secondaries at a 10 TeV Muon Collider}

\author{Ju-Yeol Choi} 
\email{joel-choi@uiowa.edu}
\thanks{\myorcid{0009-0000-7510-6377}}
\author{Matheus Hostert} 
\email{matheus-hostert@uiowa.edu}
\thanks{\myorcid{0000-0002-9584-8877}}
\affiliation{Department of Physics and Astronomy, University of Iowa, Iowa City, IA 52242, USA}
\author{Peiran Li} 
\email{li001800@umn.edu}
\thanks{\myorcid{0009-0005-7748-7085}}
\author{Zhen Liu}
\email{zliuphys@umn.edu}
\thanks{\myorcid{0000-0002-3143-1976}}
\affiliation{School of Physics and Astronomy, University of Minnesota, Minneapolis, MN 55455, USA}

\begin{abstract}
Muon decays in a muon collider ring would produce TeV neutrino and antineutrino beams of electron and muon flavor.
We study this flux in the forward $\mu^+$ and $\mu^-$ beam directions at a 10 TeV muon collider and introduce \mint, a dedicated Monte Carlo simulation to model neutrino fluxes accounting for muon beam dynamics.
We find that a benchmark detector at 5 km from the interaction point would see about $\mathcal{O}(10^{9})$ neutrino interactions per year in a $\sim3$~tonne fiducial volume with a beam spot size of $\mathcal{O}(1)$~meter.
We calculate the number of secondary muons and neutrinos generated by neutrino interactions in the rock upstream of the forward detectors and find that about two secondary high-energy and highly polarized muons from the rock would cross each detector per bunch crossing.
Neutrino productions of charmed mesons and taus in the rock generate a small $\nu_\tau+\bar\nu_\tau$ secondary flux, with $\mathcal{O}(0.2)$ events per year in the detectors, likely too small to be observed.
Wrong-sign neutrinos from similar processes, such as $\nu_e+\bar\nu_\mu$ in the $\mu^-$ beam, are more numerous but still a fraction no larger than $\mathcal{O}(10^{-8})$ of the number of TeV neutrino interactions.
Finally, we outline how the large forward neutrino exposure can be used to search for beyond-the-Standard-Model particles produced in neutrino interactions, with examples of heavy neutral leptons coupled to electron and muon flavors through mixing or electromagnetic dipole operators.
\end{abstract}

\maketitle

\setcounter{secnumdepth}{3}
\setcounter{tocdepth}{1}
\tableofcontents

\vfill

\section{Introduction}
\label{sec:introduction}

Muon colliders offer a compelling path toward the next energy and intensity frontier in particle physics.  
By combining the clean experimental environment of lepton collisions with multi-TeV center-of-mass energies, they provide powerful opportunities for precision Standard Model (SM) measurements and for searches for new physics beyond the Standard Model (BSM).  
Their physics program spans Higgs and electroweak dynamics, top-quark physics, flavor and neutrino-mass motivated scenarios, and direct probes of new particles at and above the TeV scale~\cite{AlAli:2021let,Black:2022cth,Adolphsen:2022ibf,MuonCollider:2022nsa,Hamada:2022mua,Accettura:2023ked,InternationalMuonCollider:2024jyv}.

The short lifetime of the muon requires constant replenishment of the beam and, therefore, a high-intensity proton driver and muon source.
To achieve the desired luminosity, a 10 TeV muon collider (MuC) would require $\mathcal{O}(10^{12})$ muons per bunch, with a bunch repetition rate of $\mathcal{O}(5~\mathrm{Hz})$~\cite{InternationalMuonCollider:2024jyv}, totaling $\mathcal{O}(10^{20})$ muons per year.
All muons in the MuC complex eventually decay to produce neutrinos with energies ranging from sub-GeV to multi-TeV, and
the number of neutrinos produced is comparable to that of dedicated accelerator neutrino experiments~\cite{T2K:2019eao,DUNE:2016hlj,Hyper-KamiokandeWorkingGroup:2014czz,Delahaye:2018yfq}.
Studying these neutrinos will be an integral part of the design and operation of the collider ring and detectors so as to minimize neutrino radiation hazards~\cite{King:1999ja,Mokhov:2000gt} and account for the neutrino interactions in the collider main detectors, i.e., the ``neutrino slice''~\cite{Bojorquez-Lopez:2024bsr}.

\begin{figure*}[t]
    \centering
    \includegraphics[width=\linewidth]{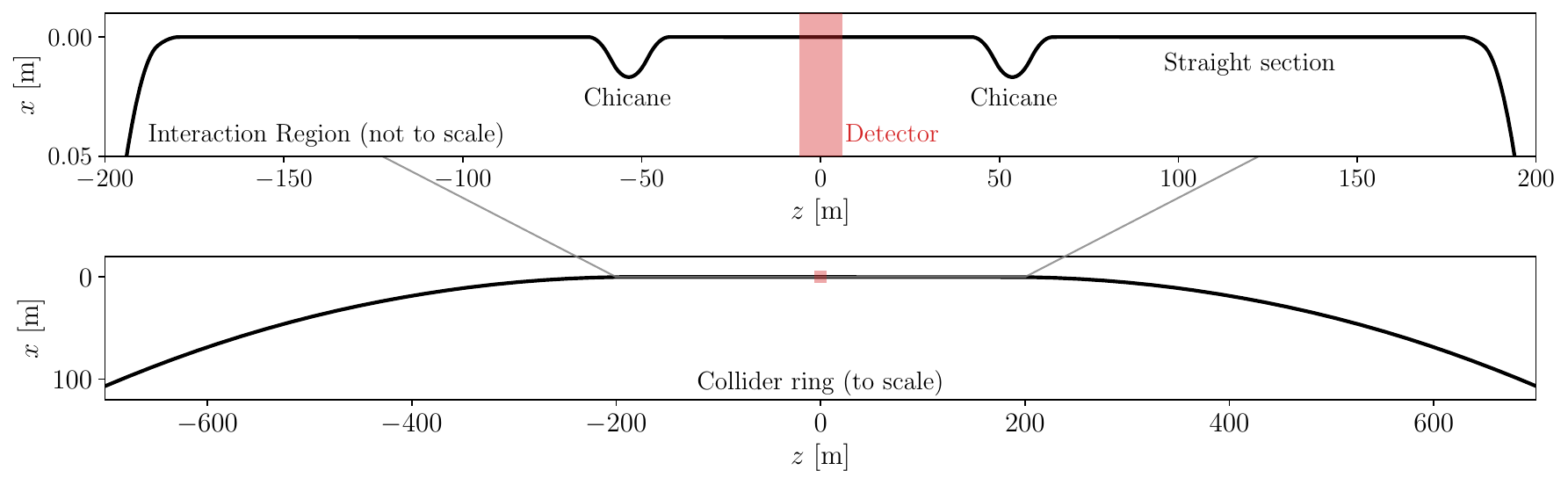}
    \caption{
    The collider ring geometry near the interaction point (0,0) of the hybrid \texttt{v0.6}+\texttt{v0.9} lattice~\cite{Vanwelde}, to scale.
    The main collider detector is shown as a red shaded region. 
    The forward neutrino facility would be located along the $x = y = 0$ direction for $z = \pm L$.
    Positive $x$ values point towards the center of the ring, while positive $y$ values point outwards.
    }
    \label{fig:ring_geometry}
\end{figure*}

Neutrinos from muon decay have long been known to provide an important complementary MuC program to study neutrino properties, electroweak and strong interactions, and search for BSM physics~\cite{Geer:1997iz,King:1997dx,Harris:1997xi,Harris:1997xj,Quigg:1997uk,King:1999kx,Bigi:2001xb,Mangano:2001mj,Huber:2014nga}.
In general, the neutrino flux from the muon beams throughout a MuC complex will be flavor-pure, highly collimated, and have a well-defined energy spectrum.
Compared with modern accelerator neutrino experiments, it is conceivable they can achieve an order-of-magnitude improvement in flux systematic uncertainties in addition to providing a unique source of high-energy \emph{electron}-flavored neutrinos.
With recent interest in a 10 TeV MuC, several new physics studies on MuC neutrinos~\cite{Adhikary:2024tvl,Liu:2024ywd,Kamp:2025yzq,InternationalMuonCollider:2025sys,deGouvea:2025zfq,Kling:2025zsb,Marzocca:2025inb}, related muon and electron beams~\cite{Cesarotti:2022ttv,Cesarotti:2023sje,Batell:2024cdl,Davoudiasl:2024fiz,Klest:2025cnd,Sakaki:2026rxj,Das:2026eyy}, and neutrino factories~\cite{Denton:2024glz,Kitano:2024kdv,Denton:2025kvy} have appeared.
The physics case for the MuC ring neutrinos has some overlap with the forward-neutrino program of FASER~\cite{FASER:2019dxq}, SND~\cite{SNDLHC:2022ihg}, and the Forward Physics Facility~\cite{Feng:2022inv,Adhikary:2024nlv} at the Large Hadron Collider, with key differences in precision, flavor content, and intensity.

In this paper, we perform a detailed study of the neutrino flux from the collider ring, focusing on the forward region of muon collisions at the interaction point (IP).
The highly collimated forward neutrinos track the beam profile of their parent muons, being sensitive to the beam optics and geometry.
Previous studies on the forward flux either neglect or simplify the beam dynamics, but, as we show, this is not a good approximation for the acceptance of a downstream detector.
In addition, we find that the energy-angle correlation (also known as the prism effect) discussed in preliminary studies for the 10 TeV MuC is washed out by the beam divergence~\cite{Calzolari:IPAC2026}.

In addition to the primary neutrino beam, we also study the flux of secondary muons and neutrinos produced by neutrino interactions in the rock upstream of the forward detector.
The secondary muon flux is particularly interesting due to its large flux, high energy, and high degree of polarization.
Neutrino secondaries from the neutrino-rock interactions are much harder to observe and likely subdominant to the component produced by muon, electron, and positron collisions at the main collider ring~\cite{Burk:2026fox}.

Finally, we explore the potential of the forward neutrino flux to probe ``secondary" BSM particles, focusing on heavy neutral leptons (HNLs) that can be produced in neutrino scattering and decay inside the detector fiducial volume.
The sheer number of interactions allows a MuC neutrino program to probe GeV-scale HNLs with mixing angles and masses beyond the reach of conventional accelerator and collider experiments.

The rest of this paper is organized as follows.  
In \cref{sec:simulation}, we discuss our simulation of the forward neutrino flux, characterize its main geometric properties, and present the resulting event-rate distributions at a downstream benchmark detector.
Then in \cref{sec:secondaries}, we calculate the flux of secondary particles produced by the forward neutrino beam in the shielding and rock upstream of the detector, including secondary muons and tau neutrinos.
In \cref{sec:new_physics}, we explore potential new physics applications of the forward neutrino flux related to new particle production in neutrino scattering, and finally,  summarize and conclude in \cref{sec:conclusions}.

\section{The Forward Neutrino Flux}
\label{sec:simulation}

\begin{figure*}[t]
    \centering
    \includegraphics[width=\linewidth]{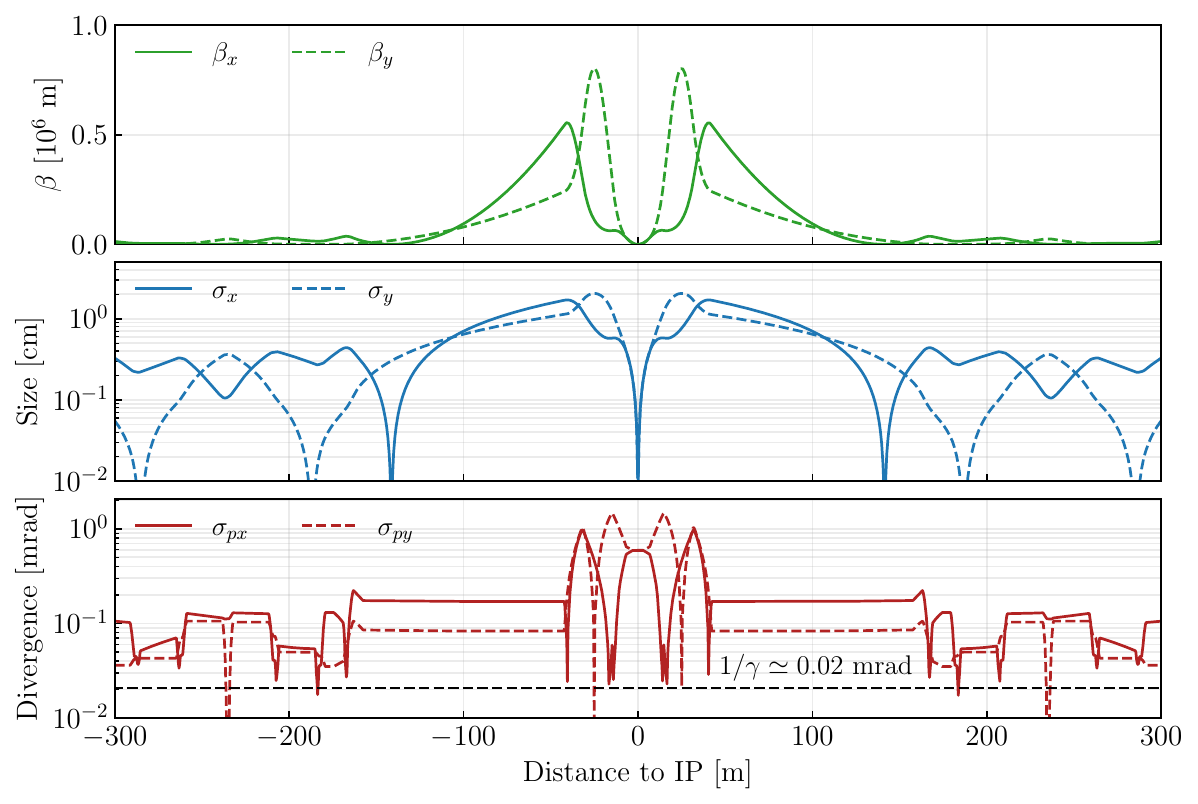}
    \caption{
    The beta function (green, top panel), transverse beam envelope size (blue, middle panel), and the beam divergence (red, bottom panel) near the interaction region of the hybrid \texttt{v0.6}+\texttt{v0.9} lattice~\cite{Vanwelde} shown in \cref{fig:ring_geometry}, assuming an RMS transverse emittance of $\varepsilon = 0.528 \times 10^{-9}$~m$\cdot$rad.
    Solid and dashed lines correspond to $x$ (horizontal) and $y$ (vertical) transverse directions, respectively.
    The typical muon-decay angular scale, $1/\gamma \sim 0.02$~mrad, is shown as a horizontal black line in the bottom panel for comparison.
    }
    \label{fig:beam_envelope}
\end{figure*}

In this section, we describe how we modeled the high-energy neutrino beam produced by muon decays around the interaction point of the 10 TeV MuC ring. 
We are primarily interested in the energy spectrum, angular distribution, and geometrical properties of the beam.
The intrinsic neutrino opening angle of the beam would be of order $1/\gamma$, but due to the parent muon beam dynamics, it is instead shaped by the properties of the parent muons and the collider geometry.
Therefore, in general, the forward-neutrino profile is affected by two main factors: i) the geometry of the ring, including the arc sections and chicanes, which change the local tangent direction of the muon trajectory away from the forward direction; ii) the transverse and longitudinal muon beam envelope and the finite muon beam divergence determined by the lattice.
A realistic prediction of the forward-neutrino flux therefore requires combining the decay kinematics with the detailed beam dynamics and geometry of the beam orbit. A complementary semi-analytical treatment of the neutrino angular distribution is provided in \cref{app:angular_distribution}.

\subsection{Muon beam simulation}
\label{sec:muon_beam_simulation}

To simulate muon decays along a realistic muon collider lattice and propagate the resulting neutrinos to a downstream detector plane, we developed a dedicated Monte Carlo called Muon Induced Neutrino Tool (\mint).
The code is publicly available through \gitlink under CC-BY-NC-SA~\cite{github}.
It simulates the trajectory of the central muon orbit and the local beam optics, allowing the finite beam envelope and beam divergence to be included consistently, but without individual particle tracking through the optics.

The input to the muon beam simulation is a Table File (\texttt{.tfs}) from a MAD-X~\cite{grote2003madx} lattice configuration.
The central orbit is constructed as short straight segments sloped towards or away from the center of the collider ring, as determined by each element of the lattice.
In this study, we use a hybrid lattice design provided by the International Muon Collider Collaboration (IMCC) that models the straight section and chicanes around the interaction region as in \texttt{v0.9}~\cite{Vanwelde:IMCC2026} and the arcs that lead up to the straight section as in \texttt{v0.6}.\footnote{The orbit-wobbling scheme proposed for neutrino-radiation mitigation applies to the main arcs, rather than to the straight section around the interaction point~\cite{Accettura:2023ked,InternationalMuonCollider:2025sys}. We therefore do not have movable beam-line elements in our forward simulation.}
The lattice accounts for about $1.5$~km of the entire ring, which we assume to have a total circumference of $10$~km.
A portion of the lattice is shown in \cref{fig:ring_geometry}.
Muon bunches are kept in the ring for $0.2$~s, corresponding to a full replacement injection rate of $5$~Hz.
About $85\%$ of the injected muons decay in our simulation.\footnote{Note that this implies the neutrino flux is decreasing with time for a single bunch, an effect that we neglect in this study.}
The duty factor of the machine is assumed to be such that one year of operation corresponds to $10^7$~s.

The muon beam dynamics is discussed in more detail in \cref{app:beam_optics}.
We assume an RMS transverse emittance of $\varepsilon = 0.528\times 10^{-9}$~m$\cdot$rad, corresponding to a normalized emittance of $\varepsilon_N = \varepsilon \, \gamma \simeq 25 \times 10^{-6}$~m$\cdot$rad.
The longitudinal spatial and momentum spread of the beams are approximated as constants throughout the ring, following a Gaussian with $\sigma_{z} \sim 1.5$~mm and $\sigma_{p_z}/p_z \sim 10^{-3}$~\cite{InternationalMuonCollider:2024jyv}.
The chicanes near the interaction point reduce the beam-induced backgrounds by deflecting electrons and positrons onto the walls before the beam reaches the detector region (see \cref{fig:ring_geometry}).
Therefore, the forward neutrino flux is dominated by muon decays along the straight section.

Muons are placed along the central orbit of the lattice and boosted to $5$~TeV. 
At each position, their transverse offsets and momentum directions relative to the local tangent of the central orbit are sampled from Gaussian distributions determined by the Twiss parameters (see \cref{app:beam_optics}).
Muons are then decayed according to the appropriate boosted decay kinematics.

Throughout our simulation, we assume the IMCC baseline of $N_\mu = 2\times10^{12}$ muons per bunch per beam, one bunch per beam, an injection frequency of $5$~Hz, and a duty factor such that one year of operation corresponds to $10^7$~s. This gives $5\times10^{7}$ injections and hence $1\times10^{20}$ muons per year in each beam. 
We keep muons in the collider ring for the full $0.2$~s and in that time, the beam covers $6.0\times10^{4}$~km, or $1.9$ muon decay lengths with $\gamma c\tau_\mu = 31$~km.
This implies that $85\%$ of the injected muons decay within the bunch storage time.
The $\pm 180$~m straight section around the IP is $3.6\%$ of the $10$~km ring, so with two neutrinos per decay and two beams it emits $1.2\times10^{19}$ neutrinos per year into the forward direction.
Of these, $7.2 \times 10^{18}$ neutrinos cross our benchmark detector face (see \cref{sec:detector_benchmark}) per year to give an acceptance of about $60\%$.

\Cref{fig:beam_envelope} shows the resulting transverse beam envelope and beam divergence around the interaction region. 
The transverse beam size is typically at the centimeter scale, whereas the beam divergence reaches the ${\cal O}(0.1-1)~\mathrm{mrad}$ level over parts of the straight section. 
This beam divergence is much larger than the intrinsic neutrino opening angle, $1/\gamma$, and therefore sets the characteristic transverse scale of the central forward-neutrino beam.

The daughter neutrinos are propagated along straight trajectories to detector planes located at $z = \pm L_{\rm det}$ of the interaction point. 
Positive $z$ values are downstream of the $\mu^+$ beam direction along the straight section (see \cref{fig:ring_geometry}).
The resulting flux is recorded as a function of the neutrino energy $E_\nu$, polar angle $\theta_\nu$, and transverse distance $r_\perp$ from the nominal beam axis.

\subsection{Detector benchmark}
\label{sec:detector_benchmark}

For our physics studies, we place identical versions of a benchmark detector upstream and downstream of the interaction region, one facing the $\mu^+$ beam and the other the $\mu^-$ beam.
We assume these detectors are placed inside a cavern that is separate from the collider ring tunnel with column density of standard rock $\rho_{\rm rock} = 2.65~\mathrm{g/cm^3}$.
The straight section ends at $z = 180$~m and the rock starts at $z = 250$~m and ends at $z = 4.99$~km, with an air gap of $10$~m between the rock and the detector face.

Because of the high energies involved, the neutrino interaction products are highly collimated along the neutrino direction, so the detector can be relatively small in transverse size.
For the same reason, good particle identification and vertex reconstruction require fine-grained tracking to resolve the highly collimated final states.
Electromagnetic and hadronic calorimeters can be placed behind the tracking system and it is desirable to have a dipole magnetic field that can separate and identify the charge of TeV muons.
To reconstruct the neutrino energy, a muon spectrometer can be placed behind the calorimeters to measure the momentum of muons that escape the calorimeter.

\begin{figure*}[t]
    \centering
    \includegraphics[width=\linewidth]{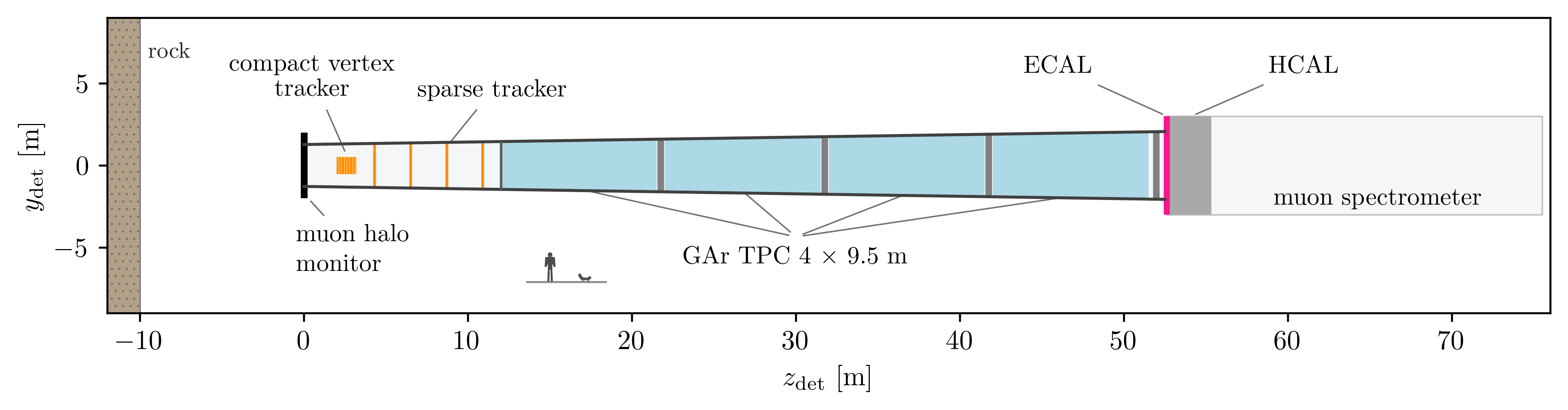}
    \caption{
    A schematic of the aspirational benchmark detector layout considered here, to scale.
    The neutrino beam enters from the left.
    The design includes a muon monitor to tag secondary muons from the rock, a compact vertex tracker ($11$~g/cm$^2$ column density) optimized for $D^\pm_{(s)}$ and $\tau^\pm$ identification with sub-mm vertex resolution followed by sparse layers to identify daughter particles, and a set of gaseous Argon TPCs adding up to $38$~m ($32$~g/cm$^2$ column density) as an active target for neutrino interactions with good particle identification. 
    Electromagnetic and hadronic calorimeters and a muon spectrometer are placed downstream.
    Both the tracker and gaseous TPC should be embedded in a dipole magnetic field for charge identification.
    Human and cat for scale.
    }
    \label{fig:detector_layout}
\end{figure*}

For concreteness, we consider the benchmark design shown in \cref{fig:detector_layout}.
Because the neutrino flux is so intense, a forward detector can trade target mass for improved particle identification and reconstruction capabilities.
This is the philosophy assumed here as well as in previous discussions~\cite{Bogacz:2022xsj,Adhikary:2024tvl,Kling:2025zsb}.
We assume the aperture to be a cone opening from $1.3$~m to $2.4$~m in radius.
The smaller radius is determined by matching the detector aperture to contain about $60\%$ of the neutrino flux from $\pm180$~m around the IP at a detector $5$~km away.
The $15$~mrad increase in aperture is assumed to contain the byproducts of neutrino interactions.

Going downstream, the detector comprises a large-area muon monitor on the front face to tag the halo of muons from rocks, a compact vertex tracker to resolve the $\mathcal{O}(1-10)$~cm decay lengths of highly-boosted charmed mesons and taus (kept deliberately narrow to catch the very center of the beam), sparse layers providing a larger lever arm to track the daughter particles, a long active tracking volume (such as high-pressure gaseous argon TPC modules), electromagnetic and hadronic calorimeters, and a muon spectrometer.
As an example, we consider four gaseous argon TPCs at $5$~bar which have a long radiation length with a total length of $40$~m, $38$ of which serves as the target for neutrino interactions with a total column density of $\rho \ell = 32~\mathrm{g/cm^2}$ on the beam axis.

\Cref{tab:detector_budget} collects the masses and on-axis column densities of the major components. 
All event rates quoted in this work are computed in the signal volume, defined as the vertex tracker plus the argon TPC, which together have a mass of $3.2$~t and an on-axis column density of $43~\mathrm{g/cm^2}$.
Note that this is a small fraction of the total detector mass, and the interactions in the pressure-vessel walls, the calorimeters, and the monitor are excluded, since a vertex cannot be reconstructed in there.
Unless otherwise specified, results in the following sections pertain to this benchmark detector.

\renewcommand{\arraystretch}{1.3}
\begin{table}[t]
\centering
\begin{tabular}{lccc}
\hline\hline
Detector component
& Length
& Mass
& Column density \\
& [m]
& [t]
& $\rho\ell$ [g/cm$^2$] \\
\hline
Muon halo monitor
& $2.0$
& $0.05$
& $0.4$ \\

Compact vertex tracker
& $1.2$
& $0.1$
& $11$ \\

Sparse tracker
& $8.8$
& $0.02$
& $0.3$ \\

Gaseous-Ar TPC
& $38$
& $3.1$
& $32$ \\

TPC walls
& $2.0$
& $4.2$
& $38$ \\

ECAL
& $0.5$
& -- 
& -- \\

HCAL
& $2.5$
& --
& -- \\

Muon spectrometer
& $20$
& --
& -- \\
\hline
Full detector
& $75$
& $470$
& -- \\
\hline
Fiducial volume
& ${39.2}$
& ${3.2}$
& ${43}$ \\[-0.2em]

\multicolumn{4}{l}{\hspace{5em}\footnotesize(vertex tracker and active TPC gas)}
\\
\hline\hline
\end{tabular}
\caption{
Lengths, masses, and on-axis column densities $\rho\ell$ of the benchmark detector components. 
The gaseous-argon TPC contains $38$~m of active gas distributed among different modules (four, in this case) within the $40$~m region from $z=12$ to $52$~m. 
The quoted $2$~m vessel length represents the cumulative thickness assigned to the internal pressure boundaries and downstream end cap, rather than a single contiguous detector region. 
Event-rate calculations in this work include only the compact vertex tracker and active TPC gas. 
Masses and column densities for the calorimeters and muon spectrometer are not used in our calculations and are therefore left unspecified.
}
\label{tab:detector_budget}
\end{table}
\subsection{Neutrino interaction rate}
\label{sec:interaction_rate}

\begin{figure*}[ht]
    \centering
    \includegraphics[width=0.46\textwidth]{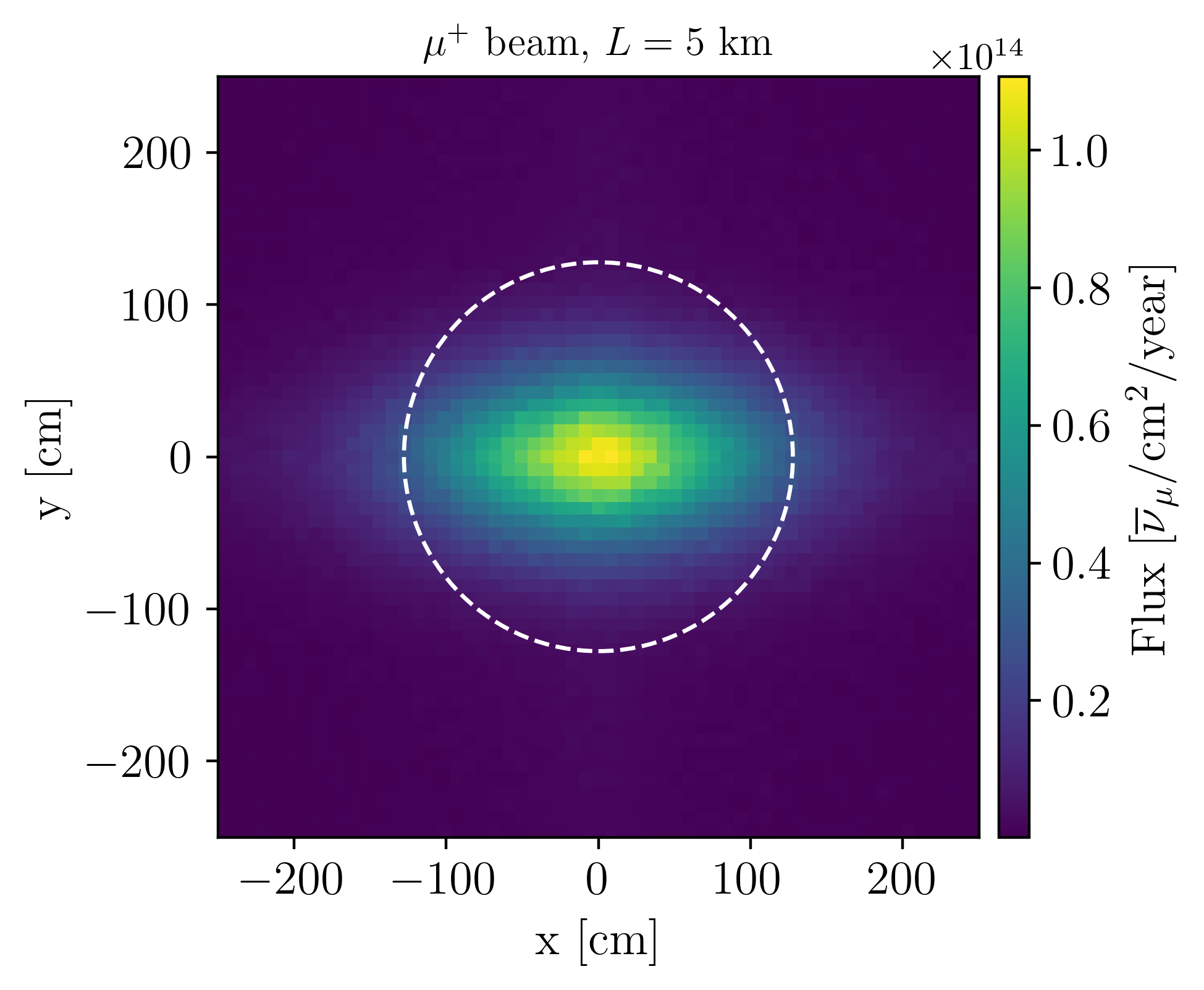}
    \includegraphics[width=0.52\textwidth]{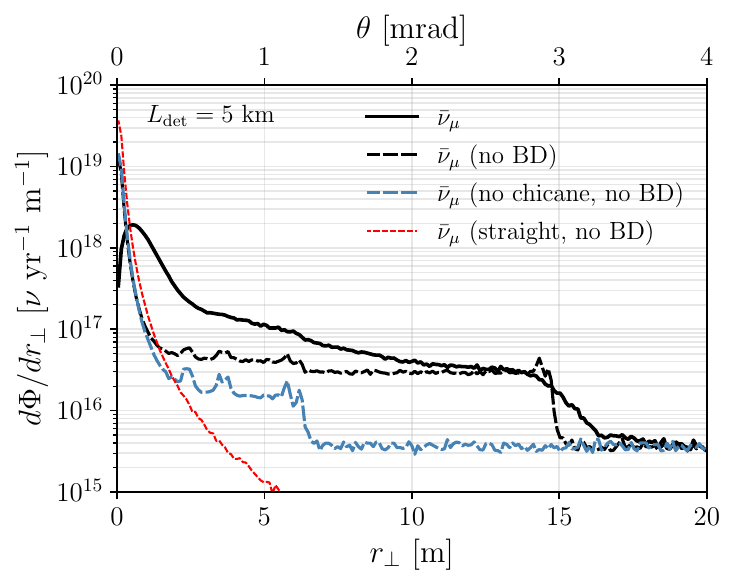}
    \caption{
    \textbf{Left:} the neutrino flux map at the detector plane for $L_{\rm det}=5$ km.
    The white dotted line represents the face of our benchmark detector.
    \textbf{Right:} the neutrino flux as a function of the radius $r_\perp$ on a detector plane located at a distance $L = 5~\mathrm{km}$ from the interaction point. 
    The different curves compare a straight lattice (dashed red), a lattice without the chicane (dashed blue), and the full lattice (black) including both the chicane and the arc sections. 
    Solid and dashed lines show the cases including and neglecting beam dynamics, respectively. 
    The beam divergence is determined by the lattice optics shown in \cref{fig:beam_envelope}.
     \label{fig:radial_and_2d_flux_map}
    }
\end{figure*}

We now present our results for the neutrino event rate at the benchmark detector discussed above. 
The left panel of \cref{fig:radial_and_2d_flux_map} shows the peak $\bar\nu_\mu$ flux at the detector to be on the order of $10^{14}$ neutrinos/cm$^2$/year on axis.
The horizontal spread is larger than the vertical spread due to the smaller beta function in the $x$ versus $y$ transverse directions of the beam (see \cref{fig:beam_envelope}).

The right panel of \cref{fig:radial_and_2d_flux_map} then plots the same flux but now per radius on the detector plane, $d\Phi / dr_\perp$, with $r_\perp = \sqrt{x_{\rm det}^2 + y_{\rm det}^2}$, for $z = \pm L_{\rm det}$.
Since the relevant angles are small, $r_\perp \simeq z\theta$, so at $L_{\rm det} = 5~\mathrm{km}$ a radius of $r_{\rm det} = 5~\mathrm{m}$ corresponds to an angle of $1~\mathrm{mrad}$.
We plot the same distribution for different variations of the lattice.
The dashed lines correspond to a simulation without beam dynamics.
We then remove the chicanes and arc sections to highlight their impact on the radial distribution.

Comparing the dashed and solid black line in \cref{fig:radial_and_2d_flux_map}, beam divergence smears the distribution, especially in the central region within $\theta \lesssim 1~\mathrm{mrad}$. 
However, the chicane and bending sections result in two plateau-like features. 
The first plateau at $r_\perp \lesssim 15~\mathrm{m}$, corresponding to $\theta \lesssim 3.0~\mathrm{mrad}$, originates from the chicanes, while the broader plateau at larger radii originates from muon decays in the arc sections, where the central orbit reaches substantially larger angles relative to the nominal beam axis.

\begin{figure}[t]
    \centering
    \includegraphics[width=0.48\textwidth]{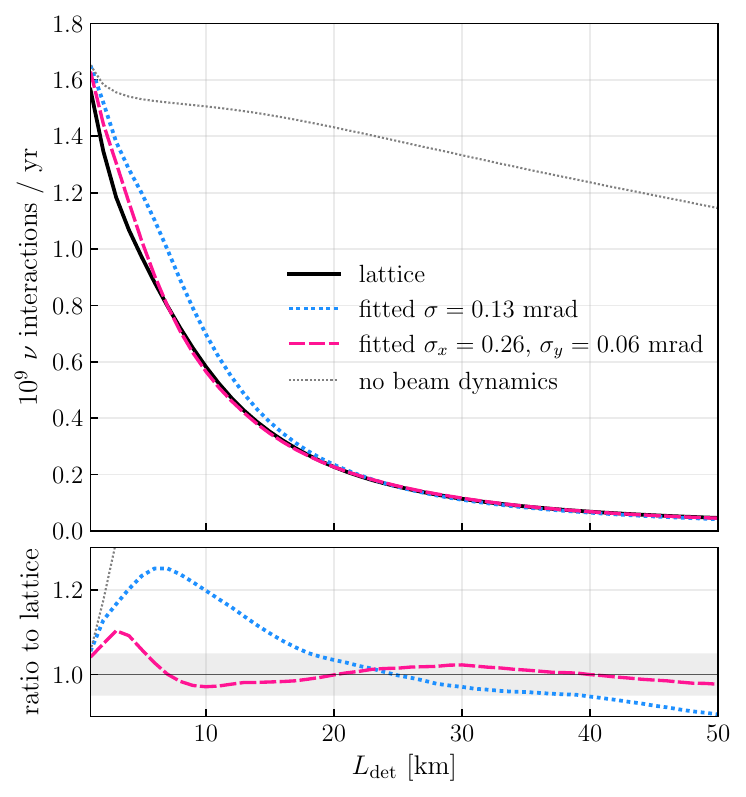}
    \caption{
    The neutrino event rate in the fiducial volume of the benchmark detector for various distances from the interaction point, $L_{\rm det} = [1, 50]$~km.
    The event rate is shown for the $\mu^+$ beam (top panel) under various assumptions about the beam divergence.
    The full lattice divergence and the no beam divergence cases are shown as a black solid line and a dotted grey line, respectively.
    We also show the results of a fit to the event rate for a constant Gaussian muon beam divergence in the $x$ and $y$ planes, taking a mock $5\%$ error band around the true lattice curve from 1 to 50 km.
    The dotted blue line shows the single-parameter fit resulting in $\sigma_{px} = \sigma_{py} = 0.13$~mrad and the dashed pink shows the two-parameter fit result with $\sigma_{px} = 0.26$~mrad and $\sigma_{py} = 0.06$~mrad.
    We find similar ratios for the $\mu^-$ beam and across flavors.
    \label{fig:rate_vs_distance}
    }
\end{figure}

\Cref{fig:rate_vs_distance} illustrates the impact of the beam dynamics on the detector acceptance by showing the neutrino event rate as a function of the distance from the interaction point, $L_{\rm det}$ for our benchmark detector.
Also shown are scenarios where the beam divergence is assumed to be a constant Gaussian, first for an isotropic scenario, $\sigma_{px} = \sigma_{py} = 0.13$~mrad, and then for anisotropic, $\sigma_{px} = 0.26$~mrad and $\sigma_{py} = 0.06$~mrad, as well as a scenario where the beam dynamics are neglected.
These values are found by fitting the neutrino interaction rate from 1 km to 50 km with a 5\% uncertainty around the true lattice curve.
The Gaussian divergence scenarios are chosen to match the typical divergence of the lattice along the largest portion of the straight section, providing a crude but useful approximation to the full lattice.

\begin{figure}[t]
    \centering
    \includegraphics[width=0.48\textwidth]{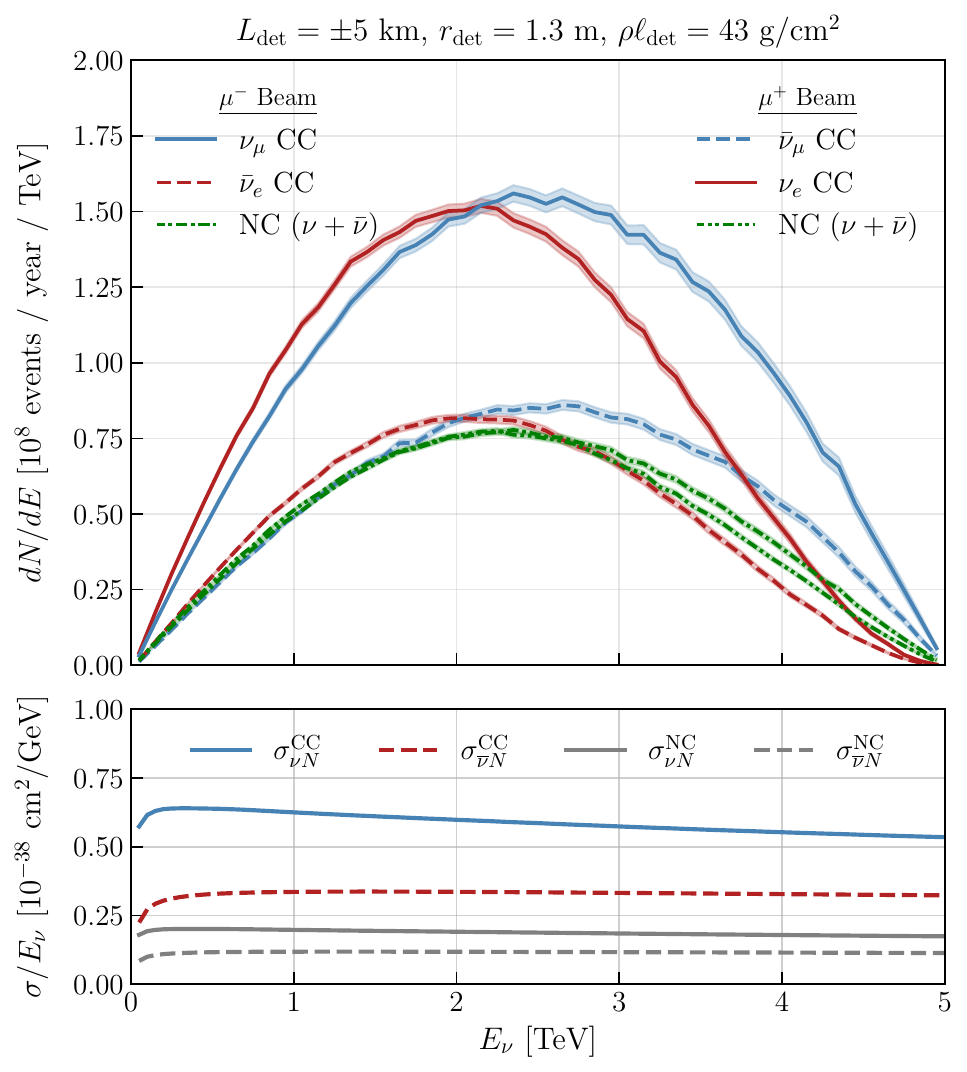}
    \caption{
    The top panel shows the neutrino energy spectrum for neutrino interactions in the benchmark detectors for the $\mu^+$ and $\mu^-$ forward neutrino beams.
    The total column density for the fiducial volume is $43$~g/cm$^2$, corresponding to the gas TPCs and the vertex tracker.
    The corresponding neutrino-nucleon cross sections for both neutrino and antineutrino CC and NC  scattering are shown in the bottom panel.
    We separate the spectra based on flavor and scattering channel: solid for neutrino CC, dashed for antineutrino CC, and dash-dotted styles for neutrino plus antineutrino NC.
    \label{fig:spectrum}
    }
\end{figure}

\Cref{fig:spectrum} shows the event spectrum of neutrino interactions for all flavors from the $\mu^+$ and $\mu^-$ beamline at identical detectors placed up and downstream of the IP.
The spectrum shows that muon flavor neutrinos will have higher energy than their electron flavor counterparts due to the muon decay spectrum.
We also see more neutrino than antineutrino events due to the larger size of the neutrino cross sections. 
The bottom panel of \cref{fig:spectrum} provides the total neutrino-nucleon cross section for CC and NC scattering as a function of neutrino energy, used to calculate the event rate at the detector.

\begin{figure*}[ht]
    \centering
    \includegraphics[width=0.49\linewidth]{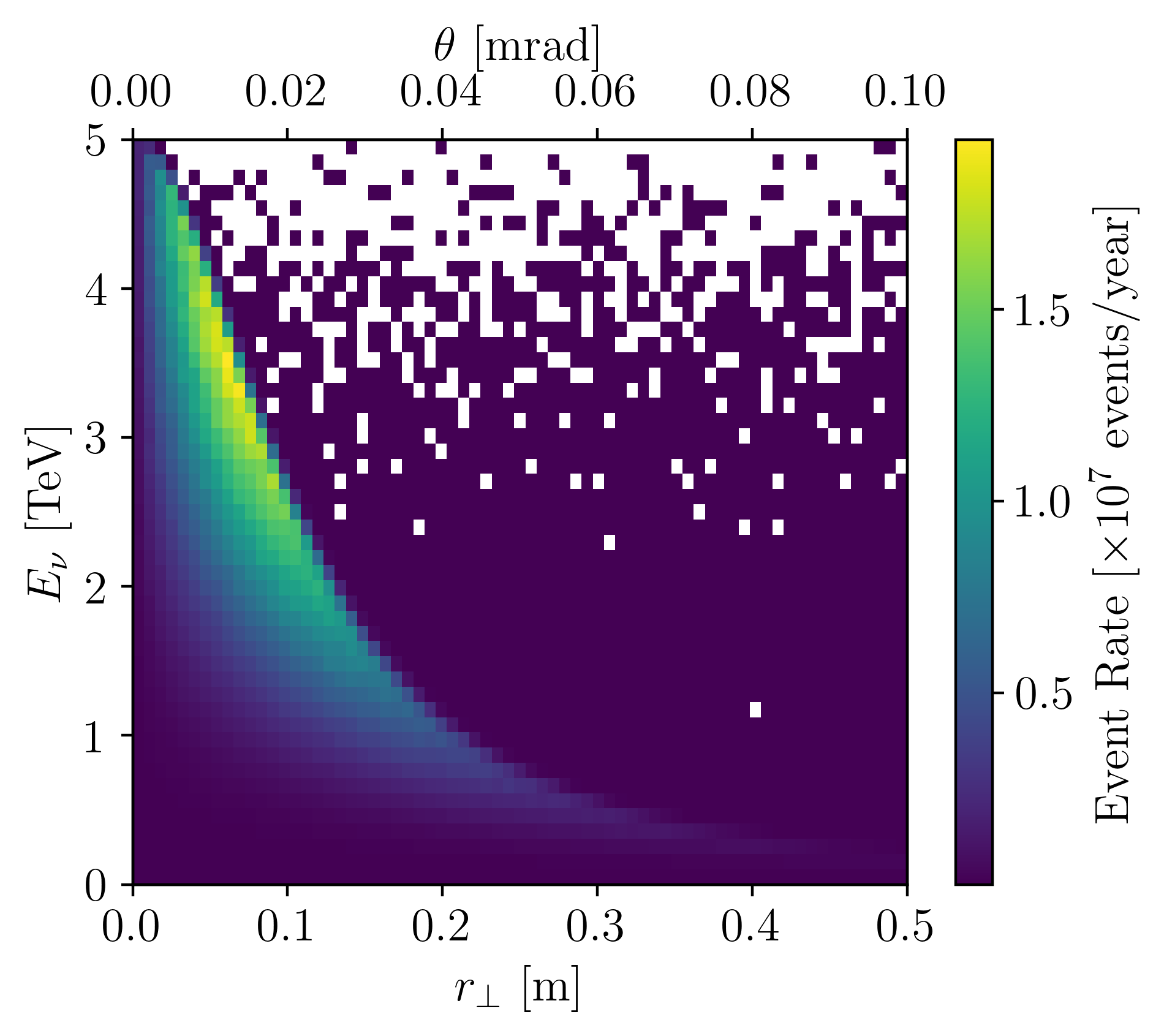}
    \includegraphics[width=0.49\linewidth]{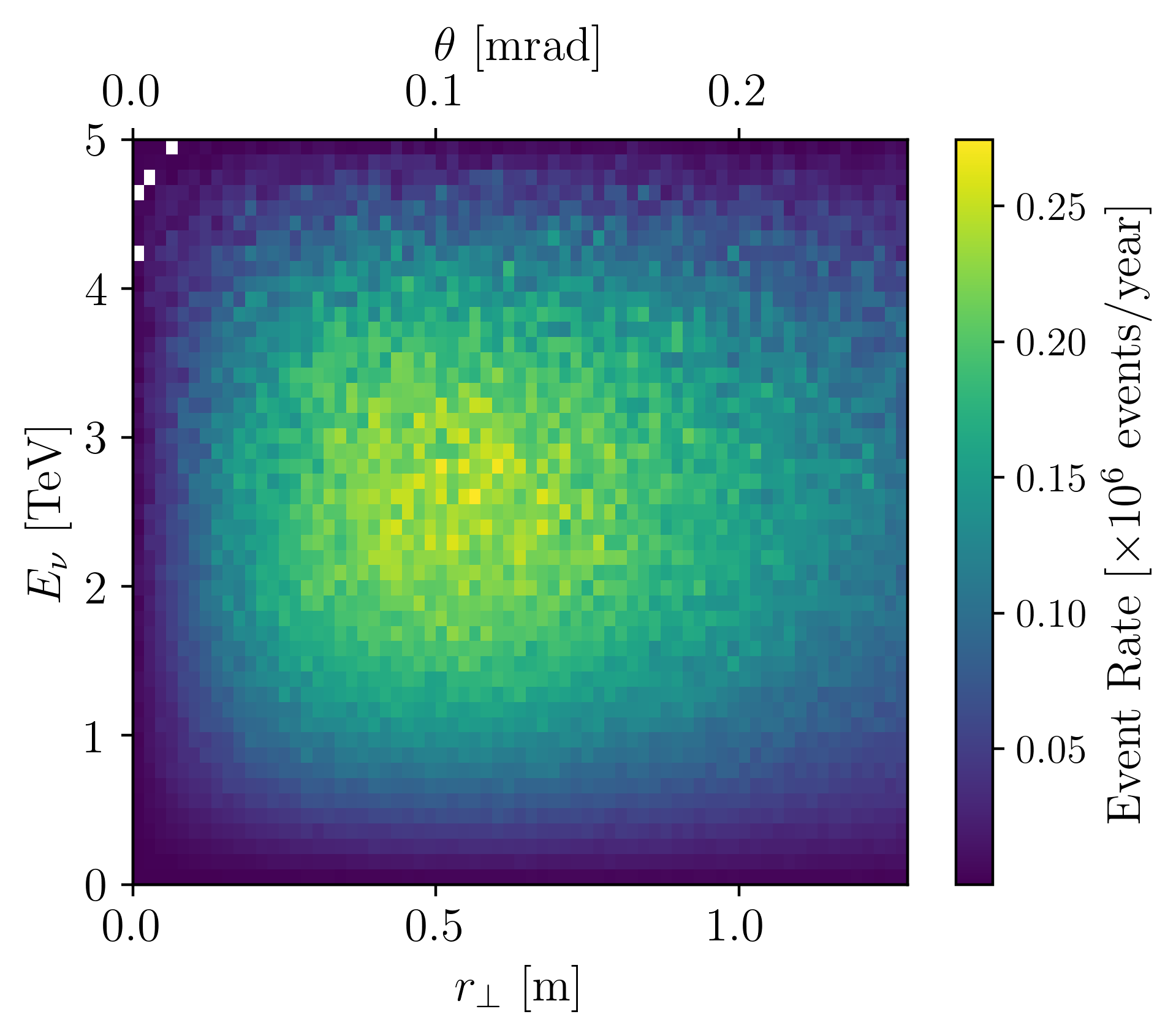}
    \caption{The two-dimensional distribution of neutrino energies and radius of interaction on the detector plane for $L = 5$~km without beam dynamics (\textbf{left}) and with beam dynamics (\textbf{right}).
    The angle of the neutrino with respect to the $\hat{z}$ direction is also shown at the top of the panel.
    Note the change in color scale and the different ranges of the $x$-axis between the two panels.
    }
    \label{fig:E_vs_radius_2D_map}
\end{figure*}

\Cref{fig:E_vs_radius_2D_map} shows the two-dimensional radial and energy distribution of the $\bar\nu_\mu$ event rate at the detector. 
Without beam dynamics (left panel), the neutrino beam is focused heavily at the center with higher energy neutrino events, clearly demonstrating a prism effect.
However, once beam dynamics are introduced (right panel), the beam is washed out to a wider spread and the neutrino energy forms a broadly peaked distribution.
Recovering the prism effect would require a much larger beta function in the straight section, such that $\sigma_{px,py} \simeq 1/\gamma_\mu = 0.02$~mrad.
For the fixed emittance of $\epsilon$, this requires $\beta$ to increase by at least a factor of $\sim 100$, implying a much larger aperture, from $5\sigma_{x,y} \sim 5$~cm to $5\sigma_{x,y} \sim 50$~cm.
We leave a feasibility study of dedicated straight sections for future work, but note that the current design is optimized for the collider luminosity and not for neutrino physics, so dedicated straight sections for neutrino and muon beam dump physics may be required.

\begin{figure*}[ht]
    \centering
    \includegraphics[width=\linewidth]{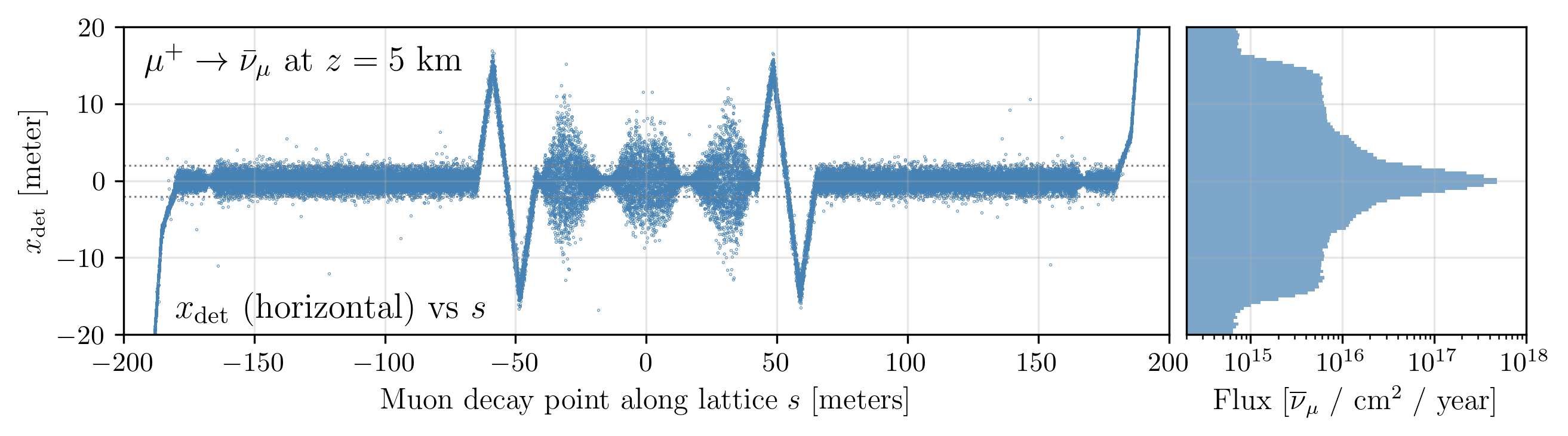}
    \includegraphics[width=\linewidth]{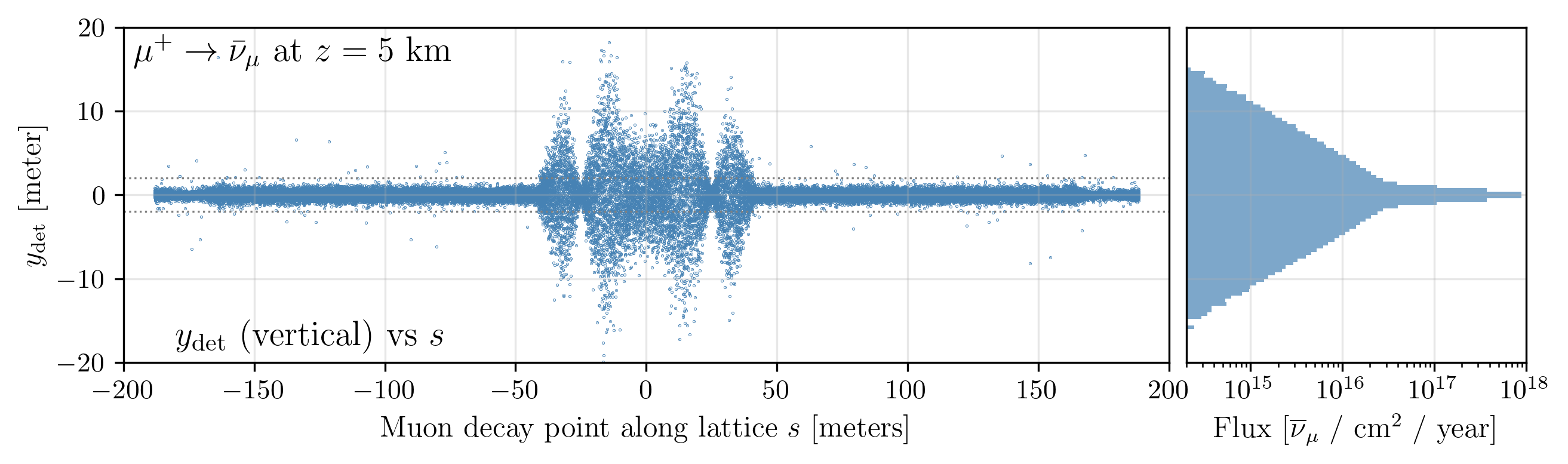}
    \caption{The neutrino production point along the lattice central orbit $s$ in the $x_{\rm det}-s$ space (top) and $y_{\rm det}-s$ space (bottom) including the beam dynamics.
    The right panels show the resulting neutrino flux as a function of the horizontal ($x_{\rm det}$, top) and vertical ($y_{\rm det}$, bottom) location of production.
    The curved sections and the chicanes are clearly visible in the $x_{\rm det
    }-s$ plane. 
    The radius of the detector is shown as a dashed grey line.
    }
    \label{fig:arrival_vs_decay_position}
\end{figure*}

\Cref{fig:arrival_vs_decay_position} plots the $x$ and $y$ coordinates of each neutrino event at the detector, referred to as $x_{\rm det}$ and $y_{\rm det}$, as a function of the location $s$ of the parent muon decayed along the lattice central orbit. 
These plots show the wider spread of neutrino events from muons near the interaction point due to the strong beam focusing.
The $x$-coordinate plot also highlights muons from the shoulders and the chicanes. 
Though the plots demonstrate we cannot clearly identify the decay location of a muon based on the transverse displacement of the corresponding neutrino event, we can more confidently say that a horizontal displacement greater than $\sim 5$~m means the neutrino most likely originated at the chicanes or arcs.

We note that the duration of the neutrino pulse from muon decays is extremely short, about $5$~picoseconds, corresponding to the longitudinal size of the muon beam, $\sigma_z = 1.5$~mm~\cite{InternationalMuonCollider:2024jyv}.
Neutrinos produced at the beginning of the straight section or at the end, arrive at the detector basically at the same time since the muon time delay with respect to a neutrino that travels the entire $360$ meters of the straight section is small:
\begin{equation}
    t_\mu - t_\nu \simeq \frac{m_\mu^2}{2 E_\mu^2} \frac{360~\text{m}}{c} \simeq 0.3 \,\text{fs}.
\end{equation}
Therefore, we conclude that neither the transverse displacement nor the timing of neutrinos can help identify their production location along the straight section in the forward direction.
This is in contrast with the neutrino slice~\cite{Bojorquez-Lopez:2024bsr} or tangential facilities~\cite{deGouvea:2025zfq} that can correlate the production point with the neutrino arrival time and direction on an event-by-event basis.

\begin{figure*}[t]
    \centering
    \includegraphics[width=\textwidth]{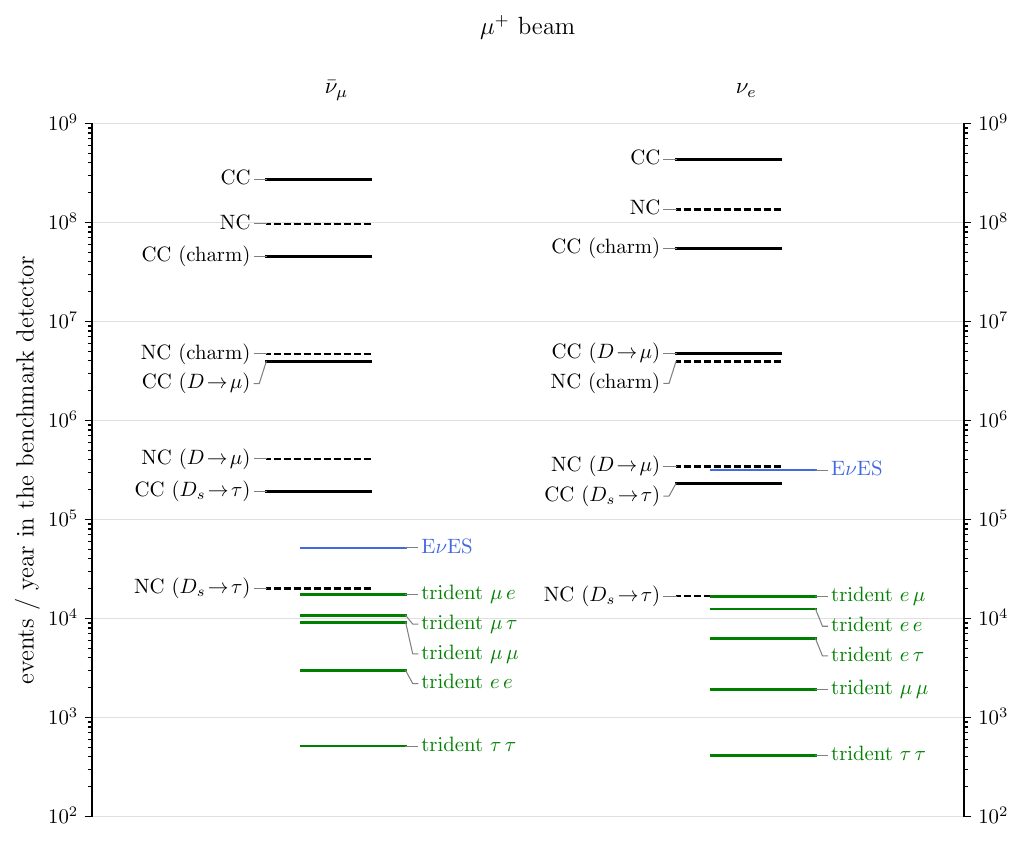}
    \caption{
    The neutrino interaction rate for various neutrino scattering channels for the $\mu^+$ beam in our benchmark detector 5 km downstream of the IP. 
    Charged-current (CC) interactions on nucleons are shown in solid lines, while neutral-current (NC) interactions on nucleons are shown in dashed lines.
    We also show the CC and NC charm production cross sections as well as the corresponding components that lead to $D \to \mu$ and $D_s \to \tau$ meson decays.
    Finally, we show the leptonic scattering channels of elastic neutrino-electron scattering (E$\nu$ES), and the several neutrino trident scattering channels including both coherent and diffractive contributions to $\nu_\alpha A \to \nu_\alpha \ell^+_\beta \ell^-_\beta A$ and $\nu_\alpha A \to \nu_\beta \ell^+_\beta \ell^-_\alpha A$.
    \label{fig:rate_summary_muplus}
    }
\end{figure*}

\begin{figure*}[t]
    \centering
    \includegraphics[width=\textwidth]{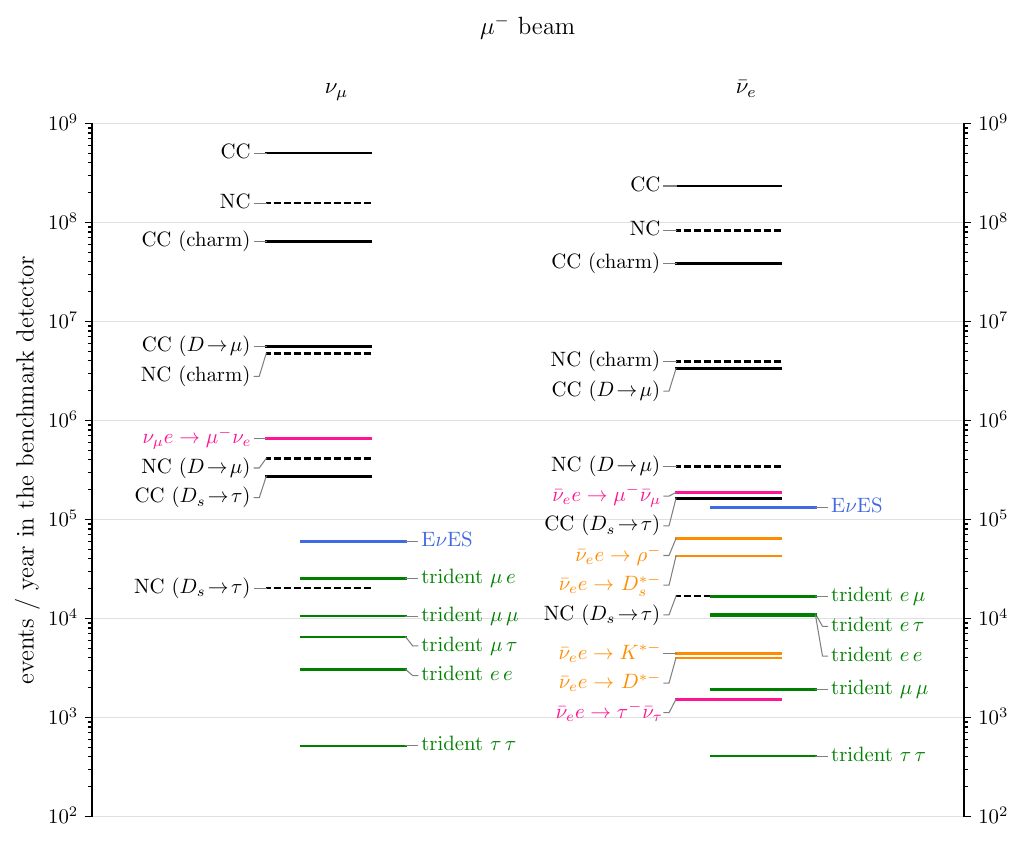}
    \caption{
    The neutrino interaction rate for various neutrino scattering channels for the $\mu^-$ beam in our benchmark detector 5 km downstream of the IP.
    Charged-current (CC) interactions on nucleons are shown in solid lines, while neutral-current (NC) interactions on nucleons are shown in dashed lines.
    We also show the CC and NC charm production cross sections as well as the corresponding components that lead to $D \to \mu$ and $D_s \to \tau$ meson decays.
    Finally, we show the leptonic scattering channels of elastic neutrino-electron scattering (E$\nu$ES), inverse muon decay (IMD), inverse tau decay (ITD), and the several neutrino trident scattering channels including both coherent and diffractive contributions to $\nu_\alpha A \to \nu_\alpha \ell^+_\beta \ell^-_\beta A$ and $\nu_\alpha A \to \nu_\beta \ell^+_\beta \ell^-_\alpha A$.    
    }
    \label{fig:rate_summary_muminus}
\end{figure*}

\Cref{fig:rate_summary_muminus,fig:rate_summary_muplus} show the energy-integrated neutrino event rates for various scattering channels, summarized for the $\mu^-$ and $\mu^+$ beams as well as different neutrino flavor separately. 
Neutrino-nucleon CC and NC interactions are dominant, with charm production providing subleading exclusive contributions relevant for muon and tau lepton production.
Leptonic scattering channels, including elastic neutrino-electron scattering (E$\nu$ES), inverse muon decay (IMD), inverse tau decay (ITD), and neutrino trident processes, are also shown.
The resonant production of vector mesons from $\bar\nu_e$ scattering on electrons are shown for $\rho^-$, $K^{*-}$, $D^{*-}$, and $D_s^{*-}$, following Ref.~\cite{Brdar:2021hpy}.
The latter resonance produces taus through the decay $D_s^{*-} \to \gamma D_s^- \to \gamma \tau^- \bar\nu_\tau$.

\section{Neutrino Secondaries}
\label{sec:secondaries}

The flux characterized above is the primary neutrino beam from muon decay in the collider ring.
A forward detector would also be exposed to a large number of secondary particles produced by neutrino interactions in the shielding, cavern wall, and rock upstream of the detector.
In this section, we estimate three secondary components that can reach the detector: muons produced mainly in CC neutrino interactions in rock, tau-flavored neutrinos produced by CC charm production, inverse tau decay, resonant vector meson production, and CC $\ell^\pm\tau^\mp$ tridents, and neutrinos with wrong signs.
Unless stated otherwise, the numbers below use the detector benchmark discussed in \cref{sec:detector_benchmark}: a detector face at $L_{\rm det}=5~\mathrm{km}$ from the interaction point, a conical aperture with starting radius of $1.3~\mathrm{m}$, a $10~\mathrm{m}$ air gap between the rock and detector face, and one operational year of exposure.

\subsection{Muon secondaries from neutrinos}
\label{sec:secondary_muons}

Muon secondaries are produced predominantly by CC interactions of primary $\nu_\mu$ or $\bar\nu_\mu$ in the last few kilometers of rock upstream of the detector.
In addition, for the $\mu^-$ beam, the $\bar\nu_e$ component can produce $\mu^-$ through IMD on electrons in the rock.
We find that IMD is a  subdominant, sub-percent contribution to the final $\mu^-$ rate.
Overall, we consider the following processes:
\begin{eqnarray}
\mu^+ \text{ beam}:& \bar\nu_\mu N\to \mu^+ X    
\\
\mu^- \text{ beam}:& \nu_\mu N\to \mu^- X, \quad \bar\nu_e e^-\to \mu^- \bar\nu_\mu.
\end{eqnarray}
While TeV muons can penetrate kilometers of rock, hadrons and electrons are absorbed within a few meters.
Nevertheless, a more realistic simulation of a forward detector environment should include the full hadronic and electromagnetic activity produced by neutrino and muon interactions in the rock, which can produce out-of-fiducial volume backgrounds in the detector.

The CC neutrino-nucleon cross section is dominated by deep inelastic scattering (DIS) at TeV energies. 
In terms of the Bjorken scaling variables $x$ and $y$, the differential cross section is~\cite{Gandhi:1998ri}
\begin{equation}
\label{eq:secondary_muon_dis_xsec}
\frac{d^2\sigma_{\rm CC}^{\nu(\bar\nu)}}{dx\,dy}
=
\frac{2G_F^2 m_N E_\nu}{\pi}
\left(\frac{M_W^2}{Q^2+M_W^2}\right)^2
{\cal Q}_{\nu(\bar\nu)} ,
\end{equation}
where $m_N$ is the nucleon mass, $M_W$ is the $W$ boson mass, $G_F$ is the Fermi constant. The quark structure functions are
\begin{equation}
\label{eq:secondary_muon_dis_kernel}
{\cal Q}_\nu=xq+x\bar q(1-y)^2,\qquad
{\cal Q}_{\bar\nu}=x\bar q+xq(1-y)^2,
\end{equation}
with parton distribution functions $q(x,Q^2)$ and $\bar q(x,Q^2)$ evaluated at the momentum transfer
\begin{equation}
Q^2=2m_NE_\nu x y.
\end{equation}
In our simulations, the $(x,y)$ distribution is sampled from a $(\log_{10}x,\,y)$ grid weighted by \cref{eq:secondary_muon_dis_xsec}, using CT18NNLO PDFs~\cite{Hou:2019efy} evaluated at $Q^2(x,y)$ for a reference $E_\nu = 1.5$~TeV.
The PDF shape is fixed at this reference energy, the $W$ propagator factor is applied event by event at the true $E_\nu$, and the rate normalization is taken from the cross-section tables of Ref.~\cite{Weigel:2024gzh}.
Individually sampling the PDF $x,y$ kinematics at fixed energies from $0.3$ to $5$~TeV range moves the mean muon energy fraction $\langle E_\mu\rangle/E_\nu$ at the percent level across the whole relevant range, well below the other uncertainties in our secondary-muon calculation.

Note that the $W$ propagator dampens the linear growth of the cross section with $E_\nu$ at about TeV energies where $E_\nu \sim M_W^2/(2m_N x y)$.
The total CC cross sections are as shown in the bottom panel of \cref{fig:spectrum}.
At $E_\nu = 2$~TeV we find the average inelasticity $\langle y\rangle_{\nu}=0.46$ and $\langle y\rangle_{\bar\nu}=0.35$.
The outgoing muon energy and angle relative to the parent neutrino are calculated as
\begin{equation}
\label{eq:secondary_muon_kinematics}
E_\mu=(1-y)E_\nu,\qquad
\theta_{\mu\nu}^2 \simeq \frac{Q^2}{E_\nu^2(1-y)} .
\end{equation}

We also simulate inverse lepton decays on the electrons of the rock, $\nu_\ell e^- \to \ell^- \nu_e$ and $\bar\nu_e e^- \to \ell^- \bar\nu_\ell$, for $\ell=\mu$ or $\tau$.
Defining $s = m_e^2 + 2 m_e E_\nu$, $y_\ell = 1 - m_\ell^2/s$, and threshold energy $E_{\nu,\text{thr}}=(m_\ell^2 - m_e^2)/(2 m_e)$, the tree-level total cross sections used in the simulation are given by~\cite{Tomalak:2022uwv}
\begin{align}
\label{eq:imd_sigma}
\sigma_{\nu_\ell}^\ell(E_\nu)
&=
\frac{G_F^2}{\pi}\frac{(s-m_\ell^2)^2}{s}\,
\Theta(s-m_\ell^2),
\\
\sigma_{\bar\nu_e}^\ell(E_\nu)
&=
\frac{2 m_e G_F^2 E_\nu}{\pi}
\bigg[
y_\ell
\!-\!
\frac{(m_e^2-m_\ell^2)(y_\ell-2)y_\ell}{4m_eE_\nu}
\nonumber\\
&\hspace{1.2cm}
+
\left(\frac{y_\ell}{3}-1\right)y_\ell^2
\bigg]\Theta(s-m_\ell^2).
\nonumber
\end{align}
The outgoing charged lepton is then sampled from the two-body kinematics and the corresponding matrix-element.
We neglect the $W$ propagator in the IMD cross section, which is a good approximation for energies much below the Glashow resonance, $E_\nu \ll M_W^2/(2 m_e) \sim 6.3~\mathrm{PeV}$.

\begin{figure}[t]
    \centering
    \includegraphics[width=0.49\textwidth]{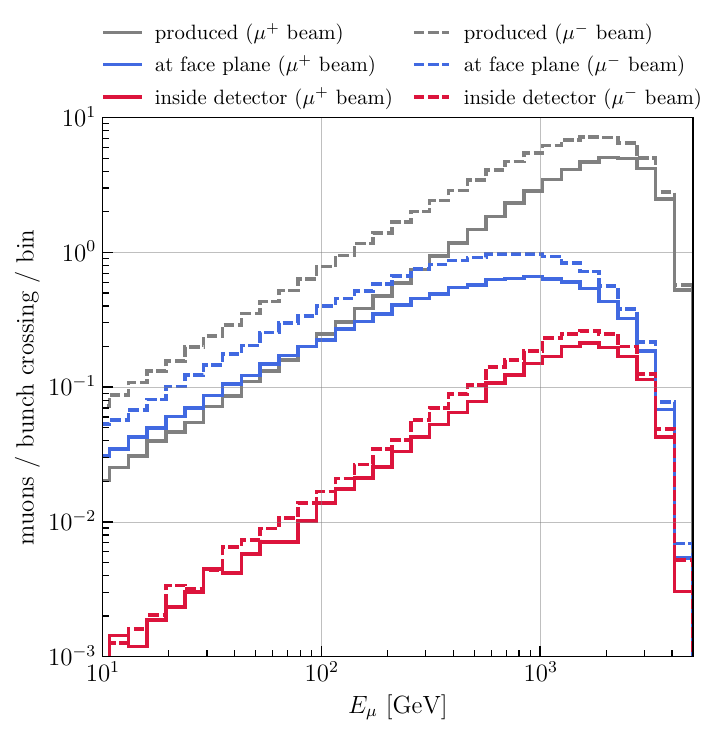}
    \caption{Energy spectrum of secondary muons produced by muon-decay neutrino interactions in the rock upstream of the benchmark detector at $L_{\rm det}=5$~km for the $\mu^+$ beam (solid lines) and the $\mu^-$ beam (dashed lines). The grey curves show the spectrum at production, the blue curves show the spectrum at the detector face after propagation, and the red curves show only the muons crossing the $1.3$~m radius detector face.
    \label{fig:secondary_muon_spectrum}
    }
\end{figure}

\cref{fig:secondary_muon_spectrum} shows the energy spectrum of secondary muons produced in the rock before and after propagation to the plane of the detector face.
We use \texttt{nuPyProp}~\cite{Garg:2022ugd} and its tables for standard rock ($\rho = 2.65~\mathrm{g/cm^3},Z_\text{eff}=11, A_\text{eff}=22$) to evolve the muon energy from the production vertex to the detector, neglecting the 10 meters of air.
The propagation includes stochastic bremsstrahlung, pair production, photonuclear losses above $y=10^{-3}$, continuous ionization and radiative losses below that threshold, and decay in flight.
The rock filters lower-energy muons (a $100$~GeV muon survives only $\sim 150$~m of rock), making the spectrum at the detector face harder than at production.

We take the muon angle with respect to the parent neutrino into account according to the deep inelastic scattering (DIS) kinematics in \cref{eq:secondary_muon_kinematics} and the IMD kinematics in \cref{eq:imd_sigma}, which are typically at the mrad scale.
\texttt{nuPyProp} propagates muons in 1D and does not include deflections of the muon trajectory, so we add multiple-scattering deflections step by step, with the angular width and correlated lateral displacement set by the Highland formula~\cite{Highland:1975pq}.
More details on the muon propagation and stochastic deflections are given in \cref{app:muon_energy_loss}.

We sample one muon production vertex per neutrino in a uniform $5$~km slab of rock upstream of the detector.
Muons produced much further upstream than $2.5$~km lose too much energy to reach the detector face, while muon production closer to the detector is suppressed by the smaller rock volume.
This also justifies the detector being located at $5$~km since at these distances, the only muons from the MuC are those produced by neutrinos, as opposed to muon beam halo losses or muon beam secondaries produced on the walls of the MuC ring.

To a very good approximation, the ultra-relativistic muons are produced as left-handed particles/right-handed antiparticles thanks to the $V-A$ character of weak interactions, with a wrong-helicity impurity at production of $\mathcal{O}(m_\mu^2/E_\mu^2)$.
Note that this is exactly the opposite of what happens in pseudoscalar meson decays thanks to the helicity flip required for angular momentum conservation.
This polarization makes the secondary muons  a strongly polarized beam, with about $\mathcal{O}(10^{12})$ muons going through the detector per year.
Stochastic energy losses (bremsstrahlung, pair production, photonuclear, and $\delta$-rays) are dominated by the chirality-preserving electromagnetic interactions with the exception of the magnetic moment nuclear interactions.
While a dedicated simulation would be needed to assert the degree of polarization lost, we estimate that the dominant depolarization is caused by multiple Coulomb scattering: it deviates the muon trajectory by $\theta_{\rm MCS}$ while the spin precesses due to the anomalous magnetic moment.
We estimate this effect to be small by folding the Highland angle over the rock column to give a flux-averaged depolarization of $\sim2\times10^{-4}$, so the muons arriving at the detector face are polarized at the level of $|P_\mu|>0.999$.

\begin{figure}[t]
    \centering
    \includegraphics[width=0.49\textwidth]{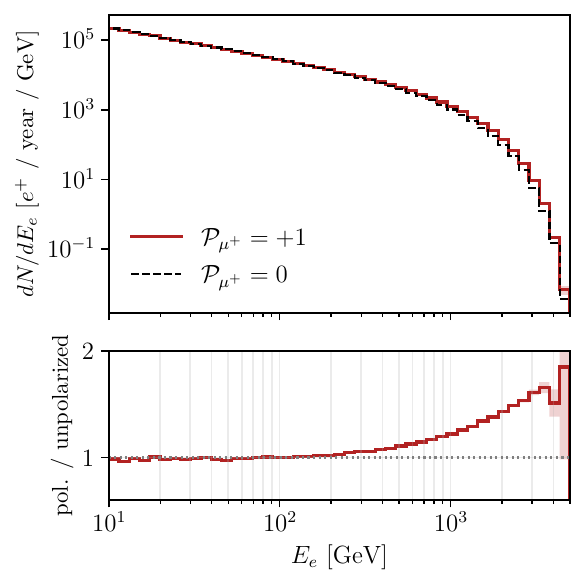}
    \caption{The energy spectrum of the positrons from secondary $\mu^+$ decays that happen between the rock and the back of the detector ($85$~m total).
    We show the expected distribution for $\simeq 100\%$ right-helical polarized antimuons in comparison with the unpolarized case for one year of operation. 
    The bottom panel shows the ratio of the polarized spectrum to ther unpolarized, illustrating how the polarization skews the high energy tail of the distribution.
    }
    \label{fig:michel_spectrum}
\end{figure}

A fraction of these muons will decay in flight to produce electrons and positrons. 
\Cref{fig:michel_spectrum} shows the resulting positron energy spectrum from the secondary $\mu^+$ in the forward direction of the $\mu^+$ beam.
We count all the decays that happen from the end of the rock layer to the back of the detector.
The muon polarization is mostly irrelevant at low energies thanks to the wide energy spectrum of the parent muons.
However, the high-energy tail is indeed sensitive to the polarization.

\begin{figure*}[t]
    \centering
    \includegraphics[width=\textwidth]{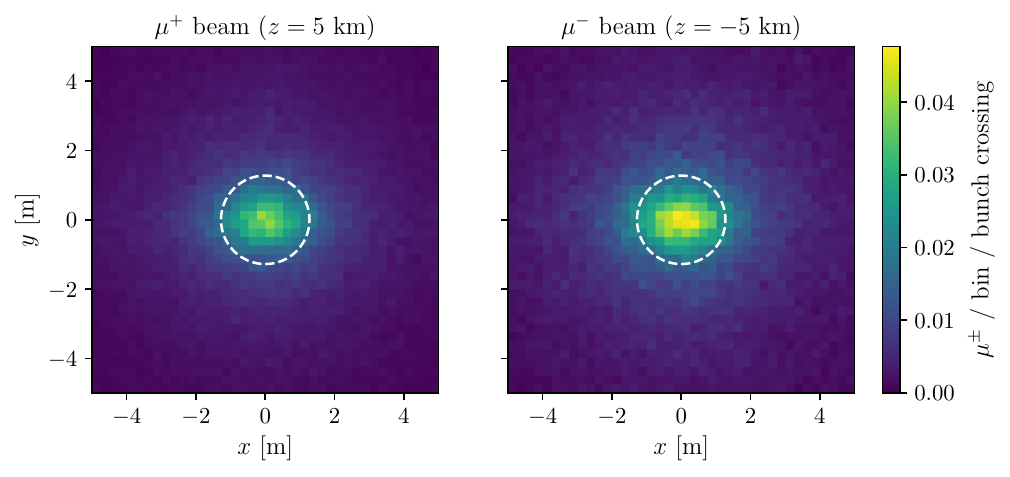}
    \caption{Transverse distribution of secondary muons at the detector face for the $\mu^+$ beam (\textbf{left}) and the $\mu^-$ beam (\textbf{right}). The white circle indicates the $1.3~\mathrm{m}$ radius of the benchmark detector.
    \label{fig:secondary_muon_face_maps}
    }
\end{figure*}

\Cref{fig:secondary_muon_face_maps} shows the secondary muon face map from the $\mu^+$ and $\mu^-$ beams at the face of our benchmark detector. 
For the $L_{\rm det}=5$~km benchmark, the primary neutrino flux through the detector face is about $3.6\times10^{18}$ $\bar\nu_\mu+\nu_\mu$ per year.
With the DIS kinematics and stochastic deflections included, the calculation gives $1.5\times10^{13}$ $\mu^+$/year and $2.5\times10^{13}$ $\mu^-$/year in $5$~km of rock.
After propagation, $6.5\times10^{11}$ $\mu^+$/year and $7.9\times10^{11}$ $\mu^-$/year cross the detector face with $\langle E_\mu\rangle  \simeq 1.2$~TeV.
With about $3.0\times10^{11}$ bunch crossings/year, this corresponds to $2.2$ $\mu^+$ and $2.6$ $\mu^-$/bunch crossing through the detector area, or $1.3$ and $1.5$ muons/cm$^2$ per second averaged over the $1.3$~m face during the $10^7$~s of operation. 
These are averages over the lifetime of the beam and, since the muon population decays away, we expect the first bunch crossings to see two times larger rates.
The IMD contribution is small, only $0.5\%$ of the accepted $\mu^-$-side rate, but is concentrated at smaller radii and higher energies than the DIS component.

The \emph{flux} of muons crossing the detector face is larger than the number of all-flavor CC+NC neutrino \emph{interactions} in a $1~\mathrm{kg/cm^2}$ detector by about a factor of $30$ for the $\mu^{+}$ beam and $35$ for the $\mu^{-}$ beam.
Secondary muons can, therefore, be a significant background for certain forward neutrino measurements.
Depending on the specific measurement, the detector design must include sufficient vetoing capabilities to reject these muons.
Note that because every bunch crossing produces a few accepted muons, vetoing on a per-bunch basis is not necessarily advantageous.
Instead, identifying and rejecting out-of-fiducial volume muons on an event-by-event basis would be preferable.
As we will see in \cref{fig:secondary_arrival_times}, the arrival time of these muons is highly overlapping with the arrival time of the neutrinos, so timing information alone is not sufficient to reject them.

\subsection{Tau-neutrino secondaries from neutrinos}
\label{sec:secondary_nutau}

Muon decays along the MuC ring produce no primary $\nu_\tau$ or $\bar\nu_\tau$. 
Interaction of TeV electrons and positron on the walls of the ring and surrounding material, however, produce a non-negligible flux of $\nu_\tau$ and $\bar\nu_\tau$ through charmed meson and tau production and decay~\cite{Burk:2026fox}.
Here, we focus only on tau neutrinos produced by primary neutrino interactions in the rock upstream of the detector.
These neutrino-induced secondaries are comparable to the photo-produced ones from the collider ring, but are produced all throughout the path of the forward neutrino beam.
Therefore, it is worth asking how this neutrino-induced secondary flux behaves as we change the detector location.
Increasing the detector distance will increase the secondary flux, but the shrinking detector acceptance to the primary neutrino beam means there's a limit to the growth.

We include four sources of secondary tau neutrinos:
\begin{enumerate}
    \item Charm production followed by leptonic decays of charmed mesons and taus,
    \begin{equation}
        \label{eq:nutau_charm_chain}
        c \to D^\pm_{(s)} \to \overset{(-)}{\nu_\tau} + \tau^\pm.
    \end{equation}
    The tau lepton subsequently decays to produce a second tau neutrino, $\tau^\pm \to \overset{(-)}{\nu_\tau} + X$.
    We neglect the smaller contribution from NC charm production~\cite{NuTeV:2000wch}.
    \item Inverse tau decay,
    \begin{equation}
        \bar\nu_e e^- \to \tau^- \bar\nu_\tau,
    \end{equation}
    giving two tau neutrinos per event, one from the primary interaction and one from the tau decay.
    \item Neutrino trident production of taus,
    \begin{equation}
        \nu_\alpha A \to \ell_\alpha^- \tau^+ \nu_\tau A, \quad \alpha = e,\mu,
    \end{equation}
    and the corresponding antineutrino processes.
    We neglect di-tau production, which has a much smaller cross section (see \cref{fig:rate_summary_muplus,fig:rate_summary_muminus}).
    \item Inverse $D_s^*$ decay,
    \begin{equation}
    \bar\nu_e e^- \to D_s^{*-} \to \gamma  (D_{s}^- \to \tau^- \bar\nu_\tau).
    \end{equation}
    The corresponding $D^*$ production can be safely neglected.
    
\end{enumerate}

Secondary tau neutrino production in the rock is a similar physical mechanism behind ultra-high-energy tau-neutrino appearance during propagation through the Earth~\cite{Soto:2021vdc}, adapted here to the TeV energies and short baselines of a muon-collider neutrino beam.
At the TeV energies of interest, $W^\pm$ and $t$-quark production are suppressed and can also be safely neglected.

\subsubsection{Charm production}
The largest secondary tau-neutrino flux is produced by CC charm production.
We account for both the $D^\pm$ and $D_s$ contributions, but the latter dominates the $\nu_\tau$ flux due to the larger branching fraction $\mathcal{B}(D_s\to\tau\nu_\tau)\simeq 5.3\%$.
The charm production cross section is suppressed either by the Cabibbo angle, $|V_{cd}|^2\simeq0.05$, or by the smaller strange-sea PDF, $s(x,Q^2)$, but it is still the dominant source of secondary tau neutrinos.

The normalization of events is obtained from the total charm production cross section, calculated with $m_c=1.70~\mathrm{GeV}$ and CT18NNLO PDFs~\cite{Hou:2019efy}.
The kinematics are sampled following a similar procedure to the dimuon charm-production treatment of Refs.~\cite{CCFR:1994ikl,Kretzer:2001tc} (see also the review in \cite{DeLellis:2004ovi}).
For neutrinos, schematically,
\begin{equation}
\begin{aligned}
\frac{d^2\sigma_c^{\nu_i}}{d\xi\,dy}
&\simeq
\frac{2G_F^2m_NE_\nu}{\pi}
\left(\frac{M_W^2}{Q^2+M_W^2}\right)^2
\\
&\quad\times
\xi\left(|V_{cs}|^2s(\xi,Q^2)+|V_{cd}|^2d(\xi,Q^2)\right)
{\cal T}_c ,
\end{aligned}
\label{eq:nutau_charm_xsec}
\end{equation}
with the slow-rescaling relation
\begin{equation}
\xi=x\left(1+\frac{m_c^2}{Q^2}\right),\qquad
{\cal T}_c=
\left[1-\frac{m_c^2}{2m_NE_\nu\xi}\right]_+ .
\label{eq:nutau_slow_rescaling}
\end{equation}
For antineutrinos, $c,s,d \to \bar c, \bar s,\bar d$.
Charm fragmentation is sampled with the Peterson function~\cite{Peterson:1982ak} with $\epsilon_P = 0.20$, and the transverse momentum about the charm jet is sampled from $dn/dp_T^2\propto\exp(-\beta p_T^2)$ with $\beta=1.21~\mathrm{GeV}^{-2}$.
The acceptance to secondary tau neutrinos is mostly dictated by this transverse momentum choice.

Overall, the $D_{(s)}$ total production rate is given by
\begin{equation}
\sigma_{D_{(s)}\to \tau \nu_\tau} = \sigma_{c/\bar{c}} \times f_{D_{(s)}^\pm} \times {\rm Br}(D_{(s)}^\pm \to\tau^\pm\nu_\tau),
\end{equation}
where: $f_{D_s^\pm}=0.08$, $f_{D^\pm}=0.26$, ${\rm Br}(D_s\to\tau\nu_\tau)=5.32\%$, and ${\rm Br}(D^\pm\to\tau\nu_\tau)=1.20\times10^{-3}$, with charm-production fractions as in Ref.~\cite{DeLellis:2002pr}.

The mesons and tau leptons decay within a few centimeters of the production vertex, so we neglect their finite decay lengths and energy losses in rock.
We sample the two-body pseudoscalar decay $D\to\tau\nu_\tau$ in the meson rest frame, including the prompt tau neutrino, and then boost the tau and neutrino to the lab frame.
Angular momentum fixes the tau helicity in the $D$ rest frame and the lab frame longitudinal polarization is tracked on an event-by-event basis from the $D$-frame decay angle, approximating $P$ by the cosine of the decay angle in the $D$ rest frame.
We define $\theta^*$ as the angle of the \emph{neutrino} momentum relative to the $D$ boost direction, so that $P_{\tau^-}=-\cos\theta^*_\nu$ for $D_s^-\to\tau^-\bar\nu_\tau$, with the sign reversed for $\tau^+$.
In terms of the tau direction, $\theta^*_\tau = \pi-\theta^*_\nu$ and $P_{\tau^-}=+\cos\theta^*_\tau$.
For a pseudoscalar decay, the $\bar\nu_\tau$ is right-helical, hence so is $\tau^-$, and in the relativistic limit, $E_\tau/E_D$ is a linear function of the decay angle.

We then sample tau decay for the most important channels.
For leptonic modes and two-body hadronic modes $h=\pi,\rho,a_1$, we use
\begin{align}
\label{eq:tau_decay_polarized}
\frac{d^2n_{\rm lep}}{dx\,d\cos\theta} &= x^2\left[(3-2x)+P(1-2x)\cos\theta\right],
\\
\frac{dn_h}{d\cos\theta} &= \frac{1}{2}\left(1+\alpha_{\nu,h}P\cos\theta\right),
\nonumber\\
&
\alpha_{\nu,h}=-\frac{m_\tau^2-2m_h^2}{m_\tau^2+2m_h^2}.
\end{align}
The leptonic spectrum gives the standard high-energy tau-decay polynomials~\cite{Gaisser:2016uoy}.
For $\tau^+$, both $P$ and the spin coefficient $\alpha_{\nu,h}$ flip sign.

\subsubsection{Inverse Tau Decay}

The cross section is given by \cref{eq:imd_sigma} with $\ell=\tau$.
At the 10 TeV MuC, only the high-energy fraction of the $\bar\nu_e$ flux is above the threshold $E_\nu>3.09~\mathrm{TeV}$.
The two-body production kinematics are sampled in the center-of-mass frame with the amplitude $|{\cal M}|^2\propto(k_{\bar\nu_e}\cdot k_{\bar\nu_\tau})(p_e\cdot p_\tau)$, followed by a polarized $\tau^-$ decay with $P=-1$.
The large Lorentz boost factor from the center-of-mass frame to the lab frame makes the products from inverse tau decay nearly collinear with the parent neutrino and therefore makes this secondary $\nu_\tau$ flux highly collimated at the detector.

\subsubsection{Neutrino Trident Tau Production}

Neutrino trident production of taus via CC is another source of secondary tau neutrinos.
Despite being suppressed by $\alpha^2$ and the multi-body phase space, trident events produce two tau neutrinos per event, one from the primary interaction and one from the tau decay.
The prompt tau neutrino is typically harder than the neutrino from the tau decay and therefore dominates the contribution to the event rate.
We simulate these events with calculations from Ref.~\cite{Ballett:2018uuc}, implemented in \texttt{NEPTUNE}~\cite{neptune}, including coherent and diffractive contributions on standard rock.
The $\tau^+$ decay is simulated with $P=+1$ using the same polarized decay treatment described above.

\begin{figure*}[t]
    \centering
    \includegraphics[width=0.49\textwidth]{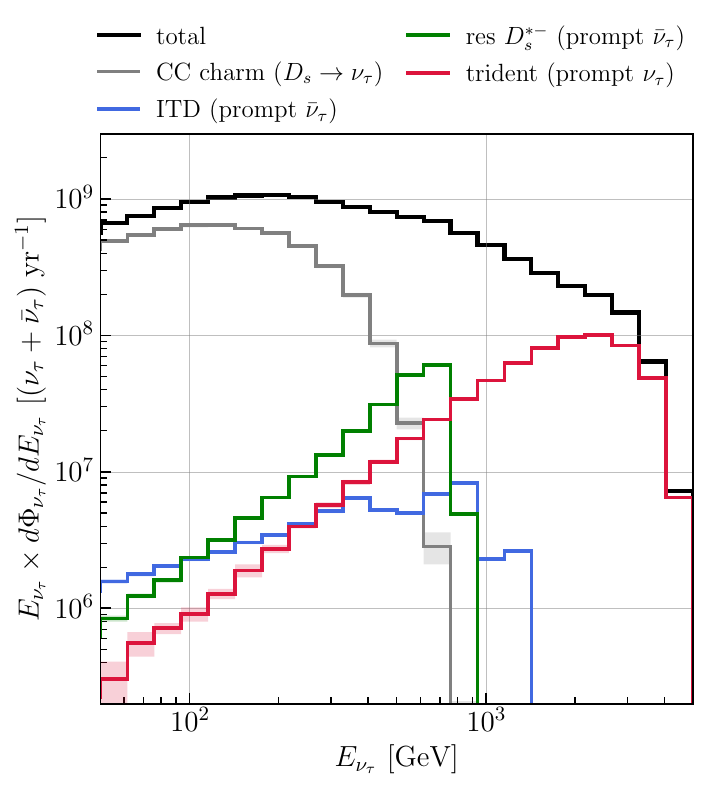}
    \includegraphics[width=0.49\textwidth]{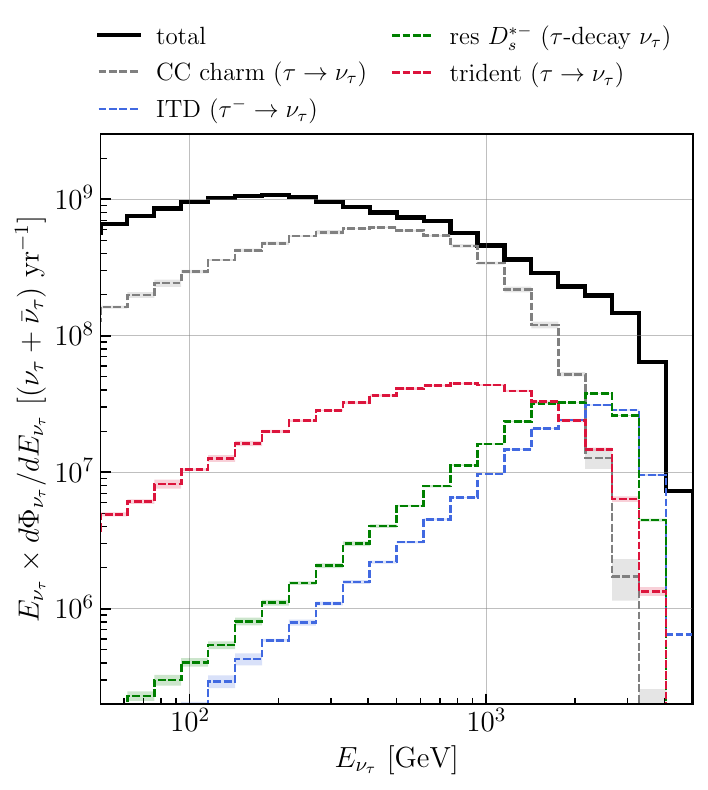}
    \caption{Secondary flux of $\nu_\tau+\bar\nu_\tau$ from muon-decay neutrino interactions in the rock through a $1.3~\mathrm{m}$-radius detector at $L_{\rm det}=\pm5$~km from the IP. 
    \textbf{Left} shows the total and the ``prompt" tau neutrino and \textbf{right} shows the total and the tau neutrinos from the decay of the tau lepton.
    We show the combined flux from the downstream and upstream directions. 
    Charm production dominates the flux, while ITD, inverse $D_s^*$ production, and CC $\ell\tau$ tridents are smaller but harder and collimated. 
    Solid lines show prompt tau neutrinos and dashed lines show tau neutrinos from subsequent tau decays.
    The shaded band shows Monte Carlo statistical uncertainties.
    \label{fig:secondary_nutau}}
\end{figure*}

\subsubsection{Inverse $D_s^*$ Decay}

Resonant production of $D_s^*$ mesons in $\bar\nu_e e^-$ scattering is yet another source of secondary $\tau$.
Following Ref.~\cite{Brdar:2021hpy}, we consider the vector meson $D_s^*$, which decays to $D_s\gamma$, followed by $D_s\to\tau\nu_\tau$.
Resonant pseudoscalar $D_s$ production is helicity suppressed and neglected.
The total cross section is
\begin{equation}
\sigma_{\bar\nu_e e^- \to D_s^{*-}}(s) = \frac{24\pi s}{m_{D_s^*}^2}\frac{\Gamma_{D_s^* \to e^- \bar\nu_e}\Gamma_{D_s^*}}{(s-m_{D_s^*}^2)^2+m_{D_s^*}^2\Gamma_{D_s^*}^2},
\end{equation}
where $s=m_e^2+2m_eE_\nu$ and $\Gamma_{D_s^*\to e^-\bar\nu_e}=G_F^2 f_{D_s^*}^2 m_{D_s^*}^3 |V_{cs}|^2/(12\pi)$ is the leptonic partial width entering the narrow-width normalization.
Numerically, we use the narrow-width approximation, for which $\int \sigma\,dE_\nu=24\pi^2\Gamma_{D_s^*\to e^-\bar\nu_e}{\rm Br}(D_s^*\to D_s\gamma)/(2m_e m_{D_s^*})$.
The $D_s^*$ is produced nearly at rest in the center-of-mass frame, and therefore the decay products are nearly collinear with the parent neutrino in the lab frame.
The subsequent $D_s$ and $\tau$ decays are simulated as above.

\begin{figure}[t]
    \centering
    \includegraphics[width=0.49\textwidth]{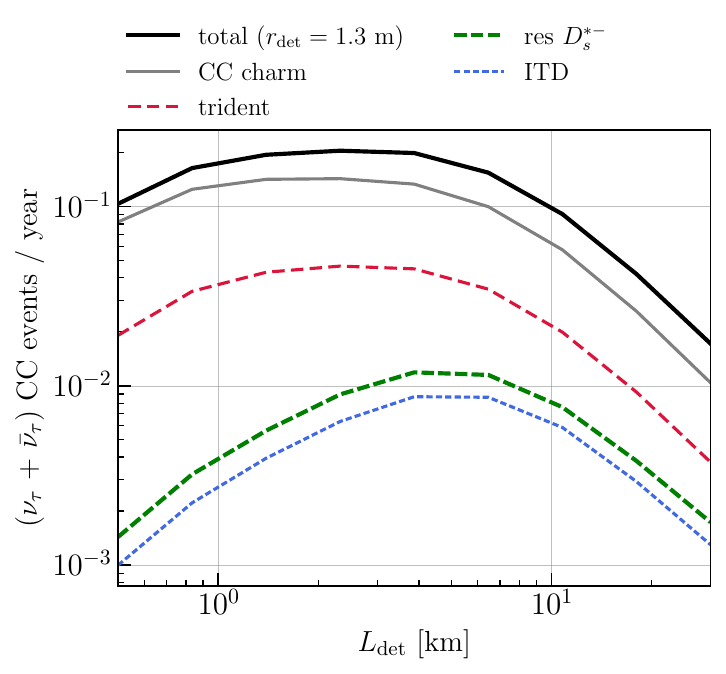}
    \caption{Secondary $\nu_\tau+\bar\nu_\tau$ CC event rate as a function of detector baseline $L_{\rm det}$ for two symmetric placements of the benchmark detector, using the fiducial volume column density $\rho\ell_{\rm det} = 43~\mathrm{g/cm^2}$. 
    The total rate has a broad maximum near $L_{\rm det}\simeq 3$~km, where the gain from the upstream rock column is balanced by the shrinking geometric acceptance of the beam. 
    These curves assume the detector is moved with a fixed geometry.    \label{fig:secondary_nutau_baseline}}
\end{figure}

At the benchmark detector, there are about $3.6\times10^{18}$ $\bar\nu_\mu+\nu_e$ per year through the detector face on the downstream side.
Before angular acceptance, these produce $1.9\times10^{12}$ $\bar\nu_\mu$-induced and $2.1\times 10^{12}$ $\nu_e$-induced charm CC events per year in the rock column, corresponding to $8.1 \times 10^9$ and $9.0\times 10^9$ $D_s\to\tau\nu_\tau$ chains per year, respectively.
After the full charm-chain kinematics and geometric acceptance, the downstream charm-induced secondary flux is $1.9\times 10^9$ $\nu_\tau+\bar\nu_\tau$ per year through the detector.
The corresponding upstream charm calculation gives a similar event contribution, while inverse $\tau$ decay adds only a percent-level correction to this number.
The resonant $D_s^*$ contribution adds $1.0\times 10^8$ $\nu_\tau+\bar\nu_\tau$ per year through the face, and CC $\ell\tau$ tridents add $2.5\times 10^8$ per year when summed over both detector directions.

\Cref{fig:secondary_nutau} shows the secondary $\nu_\tau$ flux through the detector face for the benchmark detector.
The total over the two symmetric detectors is therefore $0.2$ secondary $\nu_\tau+\bar\nu_\tau$ CC events per year at $L_{\rm det}=5$~km.
For a one-year exposure of the benchmark fiducial volume, the downstream and upstream charm components give $0.07$ $\nu_\tau+\bar\nu_\tau$ CC events each, inverse $\tau$ decay gives $0.01$ events, resonant $D_s^*$ production gives $0.015$ events, and CC $\ell\tau$ tridents give $0.04$ events.
\Cref{fig:secondary_nutau_baseline} shows how this rate changes as a function of the detector baseline.
The total rate peaks near $L_{\rm det}=3$~km with $0.2$ events per year for the benchmark detector, with charm production and tridents providing the most significant contributions.
We conclude that this rate is likely subdominant to the tau neutrino flux from photoproduction of $D_s$ mesons by electrons in the MuC ring and surrounding rock.

\begin{figure}[t]
    \centering
    \includegraphics[width=0.49\textwidth]{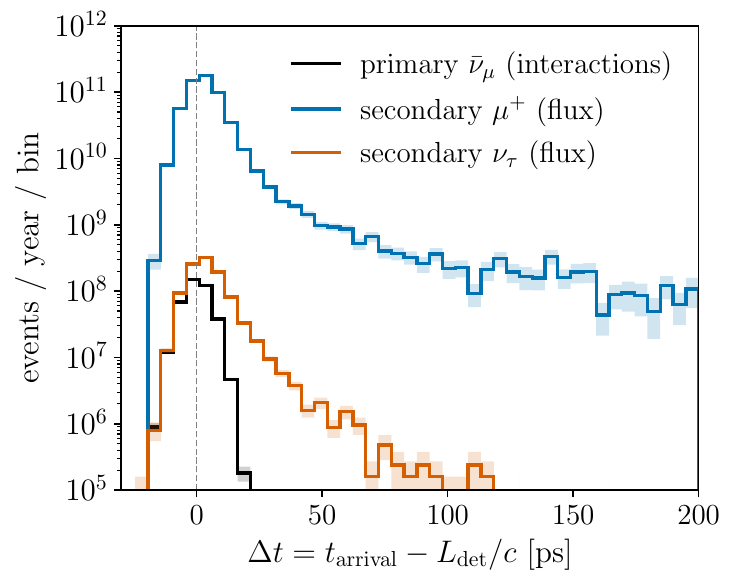}
    \caption{Arrival-time delay of accepted secondary particles relative to the primary neutrino front for the benchmark $L_{\rm det}=5$~km geometry. The distribution combines the secondary muons and tau-neutrino sources generated in the upstream rock for the $\mu^+$ and $\mu^-$ beam configurations.
    \label{fig:secondary_arrival_times}}
\end{figure}

\Cref{fig:secondary_arrival_times} shows the arrival-time distribution of secondary muons and tau neutrinos produced in the rock for the combination of $\mu^+$ and $\mu^-$ beams.
Although the muons and tau neutrinos are delayed with respect to the ps-scale bunch size of primary neutrinos from the MuC, most accepted secondaries arrive within tens of picoseconds of the primary neutrinos.
Such delays would be extremely challenging to resolve in a realistic detector, so for practical purposes, the bulk of the secondary fluxes are effectively coincident with the primary neutrino flux.

\subsection{Wrong-sign neutrinos}

Neutrino interactions can also produce a small flux of wrong-sign neutrinos, that is, neutrinos of the opposite CP to the primary beam.
For example, in the $\mu^-$ beam, the primary neutrinos are $\nu_\mu$ and $\bar\nu_e$, but secondary $\bar\nu_\mu$ and $\nu_e$ can be produced in the rock through inverse muon decay, charm production, and trident production.
We estimate these rates using the same simulation framework as for the secondary muons and tau neutrinos.
Note that muon and light meson production can also lead to wrong-sign neutrinos, but these will be predominantly low energy due to the parent particles losing energy in the rock.

\begin{figure*}[t]
    \centering
    \includegraphics[width=0.49\textwidth]{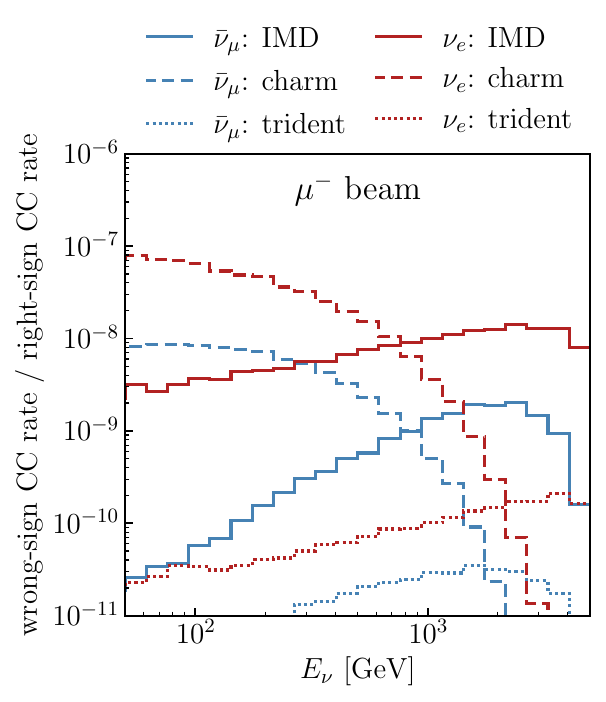}
    \includegraphics[width=0.49\textwidth]{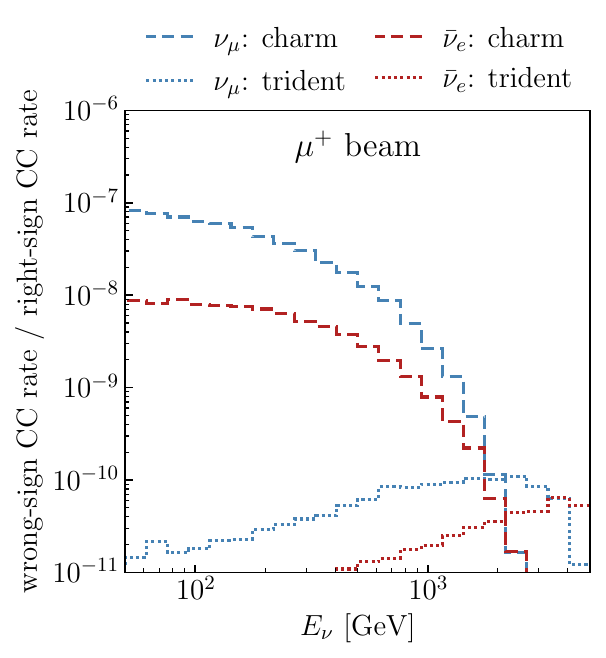}
    \caption{
        The ratio of wrong-sign (secondaries) to right-sign (primaries) neutrino CC interactions in the detector as a function of energy for the $\mu^-$ beam (\textbf{left}) and the $\mu^+$ beam (\textbf{right}). 
        Solid lines correspond to wrong-sign neutrinos produced in inverse muon decay ($\nu_\mu + e^- \to \mu^- + \nu_e$ and $\bar\nu_e + e^- \to \mu^- + \bar\nu_\mu$), dashed lines correspond to charm production followed by semi-leptonic $D$ meson decays, and dotted lines correspond to CC neutrino trident production.
    \label{fig:wrong_sign}}
\end{figure*}

For charm production, we include $D^0$, $D^\pm$ and $D_s$ followed by their inclusive semi-leptonic decays to wrong-sign neutrinos.
We use the hadronization fractions $f_{D^0}=0.60$, $f_{D^\pm}=0.26$ and $f_{D_s}=0.08$ together with the inclusive semi-leptonic branching ratios per lepton flavor, $6.5\%$, $16.1\%$ and $6.5\%$ respectively, giving an effective branching ratio of $8.6\%$ per charm CC event.
Note that these are inclusive semi-leptonic rates, but in our simulation we sample the kinematics for the dominant three-body decays,
\begin{equation}
    D^\pm \to \ell^\pm \nu_\ell + K^0, \quad
    D^\pm \to \ell^\pm \nu_\ell + K^{*0}, 
\end{equation}
which have approximately the same branching fraction for $\ell=e$ and $\ell=\mu$.
Muons and electrons are discarded, and the ``prompt" neutrino is propagated to the detector face.

\Cref{fig:wrong_sign} shows the ratio of wrong-sign to right-sign neutrino CC interactions within the benchmark detector as a function of neutrino energy for the $\mu^+$ and $\mu^-$ beams.
In terms of total rates, wrong-sign neutrinos provide at most $\mathcal{O}(6)$ CC events per year for the $\mu^-$ beam and $\mathcal{O}(1)$ for the $\mu^+$ beam in our benchmark detector.
The former is enhanced by the presence of IMD, which produces higher-energy wrong-sign neutrinos.

\section{New Physics from Neutrino Interactions}
\label{sec:new_physics}

The high intensity and collimation of the forward neutrino beam at a MuC open new opportunities for BSM searches, complementary to the forward-neutrino programs at hadron colliders and fixed-target experiments. 
The best-studied examples are precision measurements, such as the extraction of electroweak parameters from E$\nu$ES and other purely leptonic processes~\cite{deGouvea:2006hfo,deGouvea:2025zfq}, which exploit the flavor purity of the beam and the small flux systematics inherited from muon decay. 
The production of new light particles, in comparison, is less explored. 
When it comes to producing new light particles, hadronic machines like the LHC carry an apparent advantage in its large multiplicity of $pp$ collisions that give a plethora of light and heavy mesons, each opening its own production channel~\cite{Feng:2022inv}. 
At a MuC, the primary parents could be the beam muons themselves, with a kinematic reach of at most $m_\mu - m_e$.
Below this mass, however, existing constraints from low-energy muon sources and other beam dump and fixed target experiments are already quite stringent.
Instead, one can resort to particle production from $\mu^+\mu^-$ collisions, muon and secondary particle collisions with the walls, and neutrino interactions.

The forward neutrinos, together with the TeV electrons and positrons deflected onto the walls by the chicanes, deposit a large amount of energy in the shielding, tunnel walls, and rock upstream of a forward detector. 
The same interactions that generate the secondary muon and tau-neutrino fluxes of \cref{sec:secondaries} can also produce new states, with the surrounding material acting as an active target rather than as a passive absorber. 
This also removes the kinematic ceiling from muon decay. A neutrino of energy $E_\nu$ incident on a proton has 
\begin{equation}
\sqrt{s} = \sqrt{2 m_p E_\nu} \simeq 75~\GeV \left(\frac{E_\nu}{3~\TeV}\right)^{1/2}   
\end{equation}
so states with masses of tens of GeV are kinematically accessible if they can be produced in neutrino DIS. 
Because the neutrino beam is collimated and the produced states are highly boosted, the new particles continue pointing forward towards the detector with large acceptance.

The general production chain we envision consists of neutrino upscattering to a new particle, which subsequently decays to visible final states:
\begin{equation}
\label{eq:bsm_upscattering_schematic}
\nu_\alpha + T\to X_{\rm BSM} + T', \quad X_{\rm BSM} \to \text{SM particles},
\end{equation}
where $T$ is a target electron, nucleon, or nucleus in the material, and $T'$ denotes the accompanying leptonic or hadronic final state. 
If the new particle is long-lived, production inside the detector gives a displaced vertex, while production in upstream material can give an entering neutral particle that decays inside the instrumented volume. 
If the SM final states are fully visible, the event can be reconstructed as an invariant mass resonance, providing further opportunity for background discrimination.
These searches are controlled by the primary neutrino flux, the production cross section, the $X_{\rm BSM}$ decay probability, the detector response to visible final states, and the background rejection capabilities.
In the following sections, we study the first three aspects and use the resulting event yields to identify where reconstruction and background studies would be most valuable.

\subsection{Heavy neutral leptons}
\label{sec:hnl}

\begin{figure*}[t]
    \includegraphics[width=0.49\textwidth]{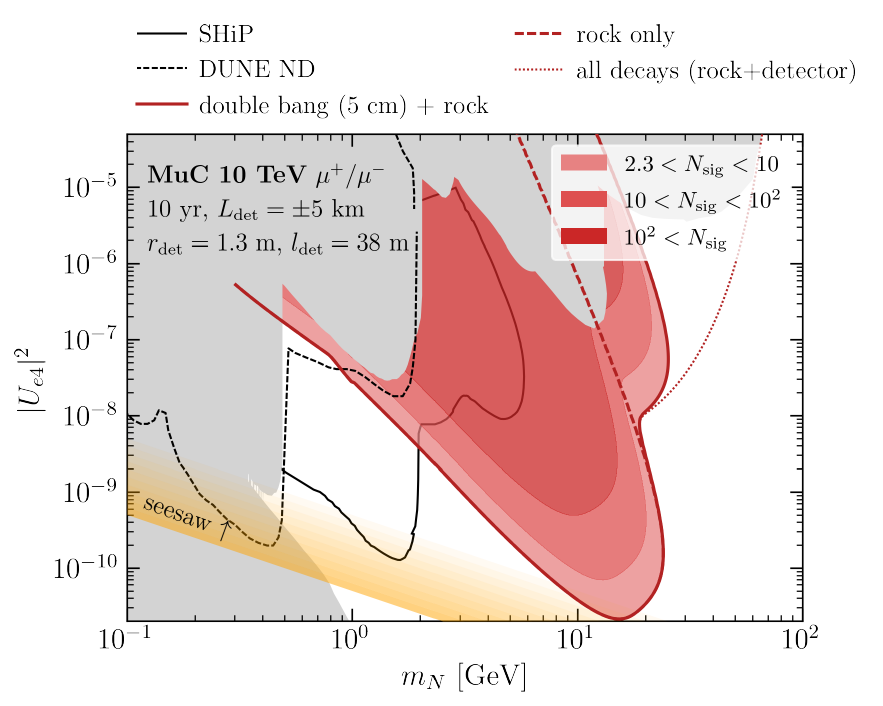}
    \includegraphics[width=0.49\textwidth]{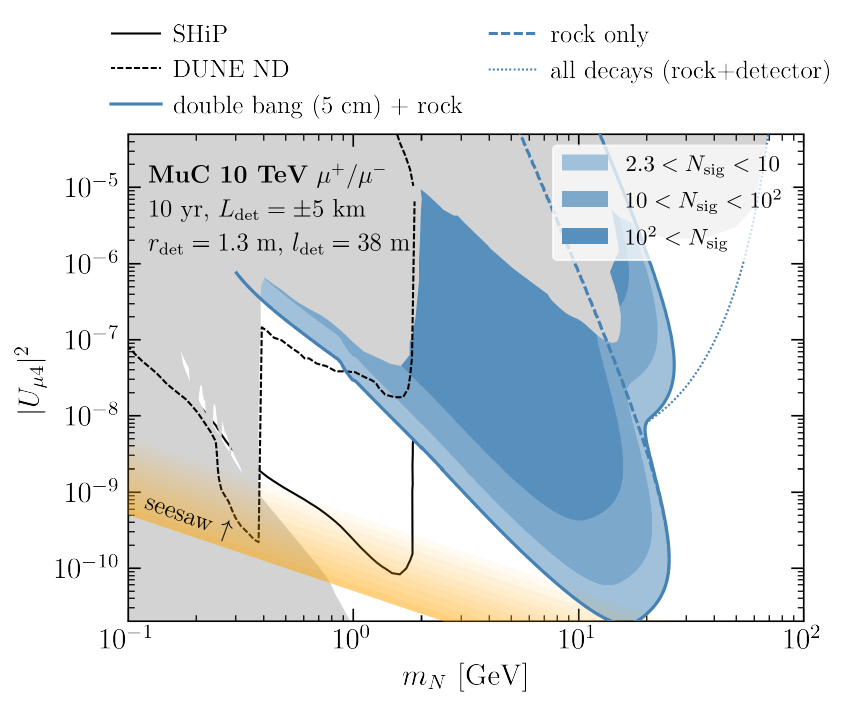}
    \caption{Estimated event-rate reach for HNLs mixed with electron (\textbf{left}) or muon (\textbf{right}) flavor, produced in neutrino interactions at a 10 TeV MuC for the benchmark detector, for a ten-year exposure of the two symmetric detectors at $L_{\rm det}=5$~km. 
    The colored curves correspond to different required signal counts after geometric acceptance but before a dedicated reconstruction and background treatment. 
    The dashed curves include rock production only, while the solid curves include both rock production and in-detector production with a visible decay displaced by at least $5$~cm from the primary neutrino interaction vertex.
    The MuC curves are event-yield regions before a full detector-background analysis.
    Shaded grey regions show the currently excluded parameter space~\cite{Fernandez-Martinez:2023phj}, while the black curves show the DUNE near detector~\cite{Berryman:2019dme} (see also ~\cite{Ballett:2019bgd,Coloma:2020lgy,Breitbach:2021gvv}) and SHiP~\cite{Albanese:2878604} sensitivities.
    \label{fig:HNL_reach}
    }
\end{figure*}

First, we consider a Majorana heavy neutral lepton (HNL) $N$ that mixes with the Standard Model neutrinos (see ~\cite{Abdullahi:2022jlv} for a review). 
We adopt a minimal single-HNL extension of the Standard Model, motivated by the type-I seesaw mechanism~\cite{Minkowski:1977sc,Yanagida:1979as,Gell-Mann:1979vob,Mohapatra:1979ia},
\begin{equation}
    \mathcal{L} \supset -\frac{1}{2} M_N \bar{N} N
    - y_\alpha \bar{L}_\alpha \tilde H N + {\rm h.c.},
\label{eq:hnl_lagrangian}
\end{equation}
where $\tilde H=i\sigma_2H^*$ and the active-sterile mixing is denoted by $U_{\alpha N}$ after electroweak symmetry breaking. 
In this minimal model, the mixing controls both production and decay. 
NC upscattering on nuclei produces $N$ in the detector or in upstream material, and $N$ decays through CC or NC weak interactions. 
Fully visible semileptonic modes such as $N\to\ell^\pm\pi^\mp$, $N\to\ell^\pm K^\mp$, and inclusive higher-multiplicity final states are especially useful because they provide a charged lepton and a reconstructable displaced vertex. 
The inclusive signal topology is
\begin{equation}
\label{eq:HNL_signature}
    \nu_\alpha + A\to N + {\rm hadrons}, \quad
    N\to \ell^\pm+X .
\end{equation}
Since the neutrino energies are in the hundreds of GeV to TeV range, the mass reach for HNLs can surpass that of the $D$ and $B$ mesons that dictate the kinematic threshold in many beam dump searches.
Indeed, this technique has already been used by high-energy experiments in the past~\cite{Mishra:1987xh,CHARMII:1994jjr} and more recently in MicroBooNE~\cite{MicroBooNE:2025khi}.
The neutrino upscattering method is also complementary to searches for HNLs produced in $\mu^+\mu^-$ collisions that target larger masses and mixings~\cite{Li:2023tbx,Mekala:2023diu,Kwok:2023dck,Cao:2024rzb}.

For the estimates shown here, the HNL production rate is computed with NC DIS as implemented in a new version of \texttt{DarkNews}~\cite{Abdullahi:2022cdw}, using CT18NNLO PDFs, $Q^2>2~\mathrm{GeV}^2$, accounting for the finite mass of $N$, and the $Z$ propagator for the benchmark detector and rock materials. Refer to \cref{app:HNL_xsec}
for upscattering cross-sections and decay channels explored for this study. The production scaling can be written as

\begin{align}
\frac{d\sigma(\nu_\alpha A \to N X)}{dx\,dy}
&= |U_{\alpha N}|^2\,
    \frac{d\sigma_Z^{\rm DIS}(E_\nu,m_N)}{dx \,dy} \label{eq:hnl_production_reweighting}
\end{align}

where $X$ denotes hadronic final states and $d\sigma_Z^{\rm DIS}(E_\nu, m_N)/dx\,dy$ denotes the differential DIS cross section including a heavy neutrino mass $m_N$ for an incoming neutrino of energy $E_\nu$.
The latter reproduces the SM NC cross section as $m_N \to 0$.
The expected number of visible decays is then
\begin{equation}
\begin{aligned}
N_{\rm sig} =
\int dE_\nu\,d\Pi\,dL\,
\Phi_\nu(E_\nu)\,
n_{A}\,
d\sigma_{\nu\to N}
\\
\times
\left( \sum_X P_{\rm dec}(E_N)\,
 d{\rm Br}_{N\to X}(m_N) \right)\,,
\end{aligned}
\label{eq:hnl_signal_yield}
\end{equation}
where $d\Pi$ denotes the production and propagation phase space,
$L$ is the length traveled by the parent neutrino through the rock, $n_A$ the number density of nuclear targets, $P_{\rm dec}$ the probability of HNL to decay, and $\text{BR}_{N\to X} (m_N)$ the differential branching ratio of the HNL to decay into a given phase space configuration of the final state $X$.
The sum is over visible final states.
The decay probability is
\begin{equation}
\label{eq:hnl_decay_probability}
P_{\rm dec}=
\exp\!\left[-\frac{L_{\rm in}}{\gamma_N\beta_N c\tau_N}\right]
\left(1-\exp\!\left[-\frac{L_{\rm det}}{\gamma_N\beta_N c\tau_N}\right]\right).
\end{equation}
with $L_{\rm in}$ the length to the face of the detector and $L_{\rm det}$ the detector length. 
The HNL widths are given in Ref.~\cite{Coloma:2020lgy}, with leptonic and exclusive meson modes below the few-GeV scale and inclusive quark-level widths at higher masses.

For in-detector production, we require the HNL decay to be displaced from the primary neutrino vertex by at least $5$~cm, and for rock production, the neutral HNL enters the detector without a visible production vertex.
For our benchmark detector, the vertex resolution would be much better than $5$~cm, so our choice is conservative.
Shrinking the required minimum displacement would enhance the sensitivity to larger mixing angles and larger HNL mass.

\Cref{fig:HNL_reach} shows the resulting signal event count contours in the mixing versus mass plane for HNLs mixing exclusively with the electron (left panel) and muon flavors (right panel).
We show our results for 10 years of a 10-TeV MuC, including both beams and both benchmark detectors place at $5$~km up and downstream of the IP.
These contours should not be interpreted as experimental sensitivity, but rather regions of interest as we do not account for detector efficiencies or backgrounds.
The $2.3$-event contour reaches down to $|U_{\mu N}|^2 \simeq 2\times10^{-11}$ near $m_N \simeq 15$~GeV, where the combination of the falling flux, the rising upscattering cross section and the requirement that the HNL decay inside the gas is most favorable.
Improving the experimental reach would be possible by considering larger detector volumes, especially in the longitudinal direction.
For in-detector upscattering, the reach is not better than about $|U_{\mu 4}|^2 \simeq 10^{-8}$, covering heavier and shorter-lived HNLs.
Considering more detector mass can help boost the sensitivity in this region.

Prompt CC and NC neutrino interactions in the detector volume can produce neutral hadrons that decay (neutral-hadron punch-through) or photons that convert cm away from the interaction vertex, mimicking the in-detector upscattering HNL signal. 
For the HNL signal from production in the rock, neutrino interactions will likely be the largest background. 
The resulting sensitivity will depend most crucially on the specifics of the detector model, its vertexing, timing, veto inefficiency, particle identification, and reconstruction capabilities.
We can readily identify several detector requirements for the HNL signatures:
\begin{itemize}
    \item an efficient upstream veto to reject entering charged particles from rock and shielding interactions,
    \item fine-grained vertexing to separate prompt neutrino interactions from displaced decays,
    \item timing resolution to associate events with a given muon bunch and reject accidental activity, and
    \item particle identification and 4-momentum reconstruction to measure the semi-leptonic invariant masses and reject neutral-hadron punch-through backgrounds.
\end{itemize}

We note that HNLs produced in the rock are delayed with respect to a speed-of-light signal from the bunch crossing.
For masses greater than $m_N \gtrsim 1$~GeV, a large fraction of signal events are delayed by as much as nanoseconds.
If such timing resolution is available, it would help mitigate neutrino backgrounds and achieve the most optimistic experimental reaches shown here.
For in-detector production, time delays will be much smaller, but the double-bang feature of the signal (first bang corresponding to scattering and the second to decay) would provide the most important discrimination power against neutrino backgrounds.

\begin{figure*}[t]
    \centering
    \includegraphics[width=0.49\textwidth]{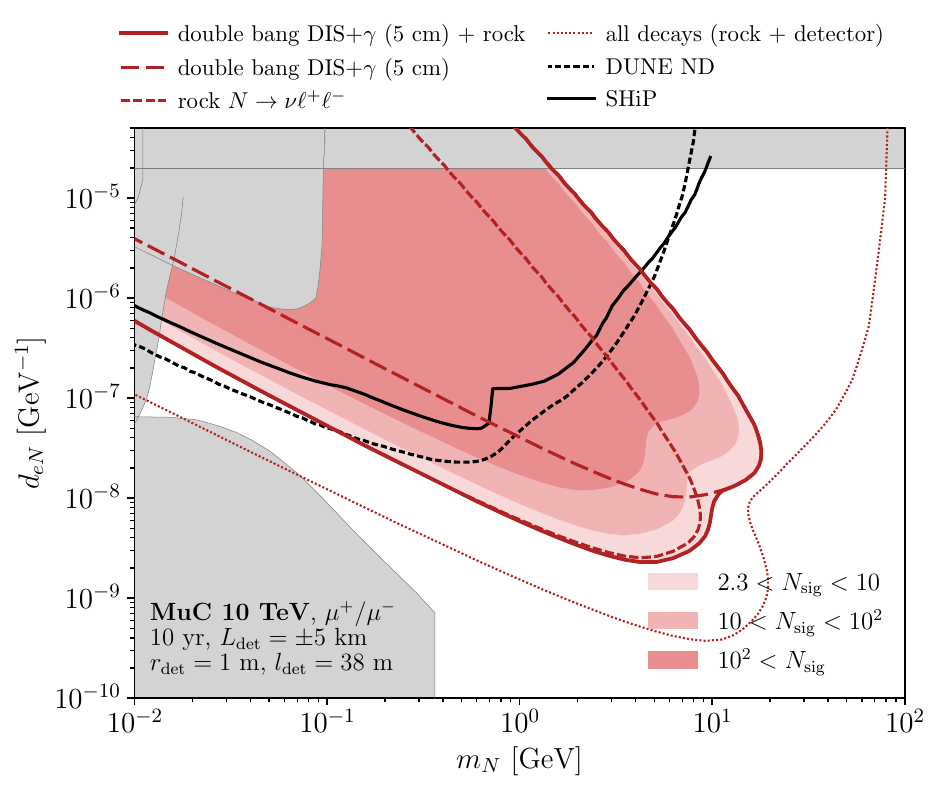}
    \includegraphics[width=0.49\textwidth]{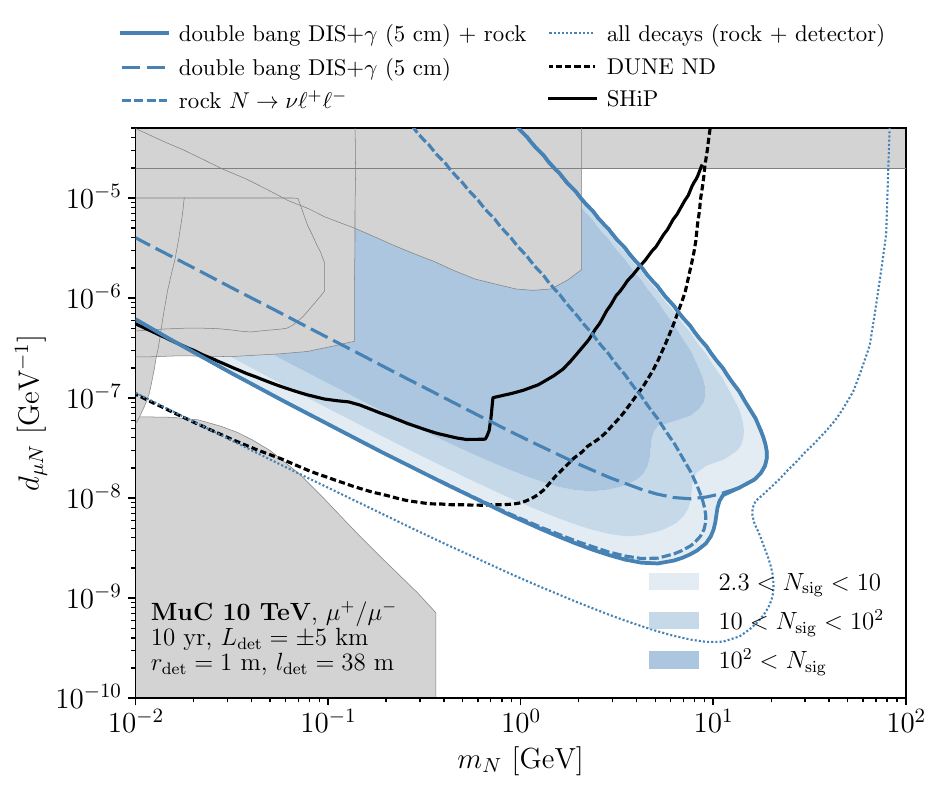}

    \caption{Estimated event-rate reach for a dipole-portal HNL at a 10 TeV MuC for the benchmark $L_{\rm det}=5$~km detector configuration, for the electron coupling $d_{eN}$ (\textbf{left}) and the muon coupling $d_{\mu N}$ (\textbf{right}), for a ten-year exposure of the two symmetric detectors. 
    The plot shows event-yield bands for a background-limited single-photon sample from coherent and detector production, a clean in-detector DIS sample with a displaced photon by at least $5$~cm, and displaced dilepton decays $N\to\nu e^+e^-$ and $N\to\nu\mu^+\mu^-$ from rock-produced HNLs. Existing limits are overlaid for comparison.
    The strongest reach is $d \simeq 4\times10^{-10}~\mathrm{GeV}^{-1}$ near $m_N \simeq 6$~GeV for the single-photon sample, and $d \simeq 3\times10^{-9}~\mathrm{GeV}^{-1}$ near $m_N \simeq 3.5$~GeV for the displaced-vertex combination.
    The MuC curves are event-yield regions before a full detector-background analysis.
    Future projects for the DUNE near detector~\cite{Schwetz:2020xra,Ovchynnikov:2022rqj} and SHiP~\cite{Magill:2018jla}, as well as existing limits in shaded grey~\cite{Magill:2018jla,Gustafson:2022rsz,Brdar:2020quo}.
    \label{fig:HNL_dipole_reach}
    }
\end{figure*}

\subsection{Dipole portal to heavy neutral leptons}
\label{sec:dipole_portal}

The minimal-mixing case is only one of many possible BSM scenarios to consider.
A useful non-minimal benchmark is a transition dipole dimension-5 operator coupling the HNL to an active neutrino, 
\begin{equation}\label{eq:dipole_portal}
    \mathcal{L} \supset
    -\frac{1}{2}M_N\bar N N
    - \left(d_{\alpha N} \bar N\sigma^{\mu\nu}\nu_\alpha F_{\mu\nu}
    +{\rm h.c.}\right),
\end{equation}
where $F_{\mu\nu}$ is the electromagnetic field strength tensor and $d_\alpha$ is the transition dipole moment with dimensions of inverse mass.
The dipole portal is a well-motivated extension of the minimal-mixing HNL, and it has been studied in the context of neutrino~\cite{Gninenko:1998nn,Gninenko:2010pr,Coloma:2017ppo,Magill:2018jla,Jodlowski:2020vhr,Brdar:2020quo,Arguelles:2021dqn,Ismail:2021dyp,Gustafson:2022rsz} and collider experiments~\cite{Ovchynnikov:2022rqj,Ovchynnikov:2023wgg,Barducci:2024kig,Beltran:2024twr,Duarte:2025zrg,Brdar:2025iua}.
At low energies, this is a dimension-5 operator.
An electroweak-gauge-invariant completion arises first at dimension six and also induces weak-boson dipoles.
At the energies of interest, the associated Feynman diagrams with $W^\pm$ and $Z$ bosons are strongly suppressed and we checked that DIS upscattering through the $Z$ boson and the CC HNL decays via the $W^\pm$ dipole are negligible compared with the electromagnetic production and decays.

Neutrino electromagnetic upscattering on nuclei produces HNLs with a rate enhanced at low momentum transfer.
For that reason, most upscattering events are coherent on the nucleus and have no visible hadronic production vertex.
For the curves in \cref{fig:HNL_dipole_reach}, we use the \texttt{DarkNews}~\cite{Abdullahi:2022cdw} coherent, proton-elastic, and DIS cross sections on the benchmark materials, with the transition-moment convention matched to the plotted $d_{\alpha N}$ coupling.
The coherent single-photon channel has the largest raw rate but is background limited, while the cleaner displaced-vertex sample combines in-detector DIS production with displaced photons and rock-produced HNLs that decay through virtual photons to $e^+e^-$ or $\mu^+\mu^-$.

The HNL decays via the dipole portal to a real photon or to a dilepton pair via a virtual photon.
\begin{equation}\label{eq:dipole_portal_signature}
    \nu_\alpha+A\to N+A,\qquad
    N\to\nu_\alpha\gamma,\qquad
    N\to\nu_\alpha\ell^+\ell^- .
\end{equation}
The radiative mode gives a single high-energy EM shower, while the dilepton mode gives a cleaner displaced vertex at the cost of a smaller branching fraction.
The decay rates can be found, for example, in Refs.~\cite{Jodlowski:2020vhr,Arguelles:2021dqn}.
Compared with the minimal HNL scenario from before, the dipole portal can have much larger production for the same lifetime, but it loses the CC decay channels that help reconstruct the HNL mass.
A realistic detector and background study is therefore even more critical and must treat photon conversion, neutral-meson backgrounds, and detector angular resolution accordingly.

The expected size of the dipole coupling is model dependent.
If it is generated by a loop of charged particles with characteristic mass $M$, coupling $g$, and chirality-breaking scale $\Lambda_\chi$, a useful parametric estimate in the convention of \cref{eq:dipole_portal} is
\begin{equation}
\begin{aligned}
\label{eq:dipole_uv_estimate}
d_{\alpha N}
&\sim
\frac{e g^2}{16\pi^2}\frac{\Lambda_\chi}{M^2}
\\
&\simeq
2.0\times10^{-8}~\mathrm{GeV}^{-1}
\left(\frac{g}{0.1}\right)^2
\left(\frac{\Lambda_\chi}{1~\mathrm{TeV}}\right)
\left(\frac{1~\mathrm{TeV}}{M}\right)^2 .
\end{aligned}
\end{equation}
This corresponds to the more common magnetic moment notation $\mu_{\rm tr}/\mu_B\simeq1.3\times10^{-10}$, using $d_{\alpha N}=\mu_{\rm tr}/2$ and $\mu_B=e/(2m_e)$.
For the 10-year, two-detector benchmark shown in \cref{fig:HNL_dipole_reach}, the background-limited single-photon event-yield target reaches its best value near $d_{\mu N}\simeq 4\times10^{-9}~\mathrm{GeV}^{-1}$ while the combined double-bang (no rock) sample reaches $d_{\mu N}\simeq 1 \times 10^{-8}~\mathrm{GeV}^{-1}$ near $m_N\simeq 7~\mathrm{GeV}$.
We also checked that a $\nu e^-\to N e^-$ with a visible electron recoil ($T_e>100~\mathrm{GeV}$) gives weaker constraints given the irreducible SM E$\nu$ES background. 
This strategy would give at best a reach of $d_{\mu N} \simeq 7 \times10^{-7}~\mathrm{GeV}^{-1}$ at $m_N \simeq 100~\mathrm{MeV}$ for the $\mu^-$-beam configuration, where the E$\nu$ES backgrounds are smallest.

Other non-minimal HNL models can be treated in the same language as above. 
For example, a dark photon or scalar coupled to the HNL can make the decay chain~\cite{Batell:2016zod,Bertuzzo:2018itn,Ballett:2018ynz,Ballett:2019pyw}
\begin{equation}
    \nu_\alpha + A \to N + A,
    \quad
    N \to \nu_\alpha A',
    \quad
    A' \to \ell^+ \ell^- ,
\label{eq:dark_photon_chain}
\end{equation}
where the dilepton invariant mass reconstructs the mediator even when the parent HNL is only partially visible. 
Direct mediator bremsstrahlung in neutrino interactions is another possibility, with emission from the neutrino, charged-lepton, electron, or quark legs depending on the couplings. 

These channels illustrate the broader point that the neutrino interaction rate around the MuC ring is large enough for the machine itself to serve as a source of weakly coupled particles.
This is especially true in the forward direction, where the straight section integrates many muon decays toward the same detector.
Because the MuC neutrinos are in the TeV range, the mass reach of this technique can outperform traditional hadronic beams when production is limited by soft hadronic exchange or by parent-meson masses.

\section{Conclusions}
\label{sec:conclusions}

The straight section around the interaction point of the 10 TeV muon collider provides an extremely high-intensity neutrino beam with unique properties.
The high energy, pure flavor composition ($\overline\nu_\mu + \nu_e$ or $\nu_\mu + \overline\nu_e$), potential precision possible on neutrino flux normalization, and strong collimation of the beam are not available to conventional hadron beam neutrino facilities. 
We have characterized this beam in detail by propagating and decaying muons along the preliminary interaction region lattice design that combines the straight section of \texttt{v0.9} with the arcs of \texttt{v0.6} of the IMCC lattice designs, taking into account the accelerator optics in \mint~\cite{github}.
We present the energy and angular distributions of the neutrino fluxes at the detector location, as well as the expected event rates for various exclusive neutrino reactions in a benchmark detector.

We considered a tentative conceptual design for a forward neutrino detector that contains both a compact tracker region for short-lived neutrino interaction byproducts and long high-pressure gaseous argon time projection chambers (TPCs) that serve both as a neutrino target and a high-resolution active volume.
This design is intended to provide a good compromise between the need for fine-grained tracking, a strong dipole magnetic field for charge identification, and a large target mass.
The signal volume used for all event rates is the vertex tracker plus the argon, with a small column density of just $43~\mathrm{g/cm^2}$ and a mass of $3.2$~t.
A lot more interactions will happen in the surrounding rock, shielding, and calorimeters, but with much less control over the event kinematics.
Identical detectors are placed at mirroring locations $\pm 5$~km away from the interaction point unless otherwise specified.

Because the muon beam needs to be strongly focused at the interaction point, the resulting neutrino flux spot size is dominated by the muon beam divergence rather than the decay kinematics, $\theta_\mu \sim \mathcal{O}(0.1~\mathrm{mrad}) \gg 1/\gamma \sim 0.02~\mathrm{mrad}$, washing out neutrino energy-angle correlations (the so-called prism effect).
Nevertheless, the straight section provides a uniquely intense and collimated neutrino flux at the MuC complex, with a total of about $7\times 10^{18}$ $\nu + \bar\nu$ per year passing through the benchmark detector face, summed over the four flavors and the two beams.
This corresponds to about $\mathcal{O}(10^{9})$ neutrino-nucleus interactions per year in the low-density fiducial volume, with about $5 \times 10^{5}$ neutrino-electron scattering events.

We also studied three examples of secondary fluxes produced by neutrino interactions in the rock: penetrating muons from neutrino CC scattering, tau neutrinos from neutrino production of charmed mesons and taus, and wrong-sign neutrinos (e.g., $\nu_\mu+\bar\nu_e$ fluxes in the $\mu^+$ beam) produced in right-sign neutrino interactions.
The secondary-muon flux at the benchmark detector face has $\langle E_\mu \rangle \simeq 1.2$~TeV after propagation and is produced predominantly in the last kilometer of rock before the detector hall.
At $L_{\rm det}=5$~km, we find about $\mathcal{O}(10^{12})$ accepted $\mu^\pm/\mathrm{year}$ through the detector face on average, corresponding to about $2$ penetrating muons per bunch crossing in each detector.
This flux is dominated by DIS $\nu_\mu$ and $\bar\nu_\mu$ CC interactions, while
IMD contributes only $0.6\%$ of the accepted $\mu^-$-side with more central and higher energy events.

We note that the secondary muons would be strongly polarized as left-helical muons/right-helical anti-muons, and could be used as a fixed target experiment with a wide-band polarized TeV muon beam.
The polarization is inherited from the CC production vertex and we expect propagation through the rock to degrade it by a negligible level.
For comparison, the secondary muon flux is about a hundred times smaller than at the high-energy polarized muon beam at the CERN COMPASS experiment where $P_\mu \simeq -0.8$ and $E_\mu \simeq 160 - 200$~GeV~\cite{COMPASS:2007rjf}, and than the $E_\mu \simeq 470 - 490$~GeV beam at the E665 experiment~\cite{E665:1989lkq}.
However, it is born with a larger polarization of the opposite sign, $P_\mu \simeq -1$ for $\mu^-$ and $P_\mu \simeq +1$ for $\mu^+$, and with a much higher energy of $\langle E_\mu \rangle \simeq 1.2$~TeV.

Secondary $\nu_\tau+\bar\nu_\tau$ fluxes from neutrino interactions are also produced, but in likely unobservable rates.
We find $0.2$ tau-neutrino CC events per year across the two symmetric benchmark detectors at $L_{\rm det}=5$~km.
Charm production is the dominant contribution, CC $\nu_\ell \to \ell^\pm\nu_\tau\tau^\mp$ tridents are the second largest, and inverse tau decay and inverse $D_s^*$ production provide smaller but higher-energy and more collimated components.
By moving the benchmark detector, we find that the total secondary tau-neutrino rate peaks near $L_{\rm det}\simeq 3$~km, falling steeply after due to the decreasing detector acceptance.
This flux does not include muon and electron collisions at the main collider ring~\cite{Burk:2026fox}.

We also explore the potential of the large forward neutrino interaction rate as a source of new particles.
In this study, we considered HNLs coupled to the SM through neutrino mixing and through effective dipole operators.
The sensitivity of a forward neutrino detector to these new particles will ultimately be limited by the ability to reject neutrino interaction backgrounds in and around the detector and will depend very strongly on the detector design.
For that reason, we present event-yield estimates over a 10-year exposure, identifying regions of interest for future detector studies.
We find large regions of open parameter space where a forward neutrino detector at a muon collider could produce $\mathcal{O}(2.3-10^3)$ visible HNL decays over this exposure.
A background-free search is unlikely, but these examples illustrate how the forward direction of a MuC can provide enough ``neutrinos-on-target'' to make neutrino upscattering a competitive probe of new physics, especially above the mass range accessible to meson-decay beam dumps.

\emph{Note Added:} In the final stages of our manuscript preparation we learned about independent work by F.~M.~Burk, T.~Han, W.~Kilian, F.~Kling, J.~Kopp, and
Z.~Tabrizi~\cite{Burk:2026fox}. 
We explore different aspects of forward neutrino flux at muon colliders. Our results are compatible and complementary.

\section{Acknowledgements}
We thank Marion Vanwelde for supplying the collider lattices used throughout this work.
We also thank Innes Bigaran, Paddy Fox, Mary Hall Reno, Tova Holmes, Daisy Kalra, Kevin Kelly, Larry Lee, and Pedro Machado for discussions on this topic.
We acknowledge the use of Claude Code and OpenAI Codex for coding and text editing. 

P.L. and Z.L. are supported by the Department of Energy under Grant No.~DE-SC0011842 at the University of Minnesota. P.L. is partly supported by a Doctoral Dissertation Fellowship at the University of Minnesota. Z.L. is supported in part by a Sloan Research Fellowship from the Alfred P. Sloan Foundation at the University of Minnesota. This work was performed in part at Aspen Center for Physics, which is supported by National Science Foundation grant PHY-2210452.
M.H. was partially supported by the University of Iowa's Year 2 P3 Strategic Initiatives Program through funding received for the project entitled ``High Impact Hiring Initiative (HIHI): A Program to Strategically Recruit and Retain Talented Faculty.''

\appendix

\section{Neutrino Angular Distribution}
\label{app:angular_distribution}
 
It is useful to derive the angular distribution of the neutrino events from boosted muon decays observed at a detector located $z=L_{\rm det}$ away to compare with our \mint results. 
We neglect neutrino masses and assume the muon decays are unpolarized, $P_\mu = 0$.
A non-zero $P_\mu$ would tilt the decay spectrum, changing the $\nu_e$ and $\nu_\mu$ energy distributions in opposite directions while leaving their sum nearly unchanged.
In the $P_\mu = 0$ limit, after integrating over the neutrino energy, the neutrino direction is isotropic in the muon rest frame.

First, consider a parent muon moving along the positive $z$ direction in the lab frame, with velocity $\beta$ and boost factor $\gamma = E_\mu/m_\mu$.
In the muon rest frame (denoted by a star) the neutrino four-momentum
\begin{align}
    (p^*_\nu)
    =
    E^*_\nu
    \left(
    1,\,
    \sin\theta^*_\nu \cos\phi^*_\nu,\,
    \sin\theta^*_\nu \sin\phi^*_\nu,\,
    \cos\theta^*_\nu
    \right),
\end{align}
gives the lab-frame components after a boost of $\beta$ along the $z$ direction:
\begin{align}
    E_\nu
    &=
    \gamma E^*_\nu \left(1+\beta\cos\theta^*_\nu\right),
    \\
    p_{\nu,z}
    &=
    \gamma E^*_\nu \left(\cos\theta^*_\nu+\beta\right),
    \\
    \vect{p}_{\nu,\perp}
    &=
    E^*_\nu \sin\theta^*_\nu
    \left(
    \cos\phi^*_\nu,\,
    \sin\phi^*_\nu
    \right).
\end{align}
The lab-frame polar angle $\theta$ therefore satisfies
\begin{align}
    \cos\theta_\nu
    =
    \frac{p_{\nu,z}}{E_\nu}
    =
    \frac{\cos\theta^*_\nu+\beta}
    {1+\beta\cos\theta^*_\nu},
    \label{eq:app_aberration_forward}
\end{align}
or equivalently,
\begin{align}
    \cos\theta^*_\nu
    =
    \frac{\cos\theta_\nu-\beta}
    {1-\beta\cos\theta_\nu}.
    \label{eq:app_aberration_inverse}
\end{align}
The solid-angle Jacobian gives
\begin{align}
    d\Omega^*_\nu
    =
    \frac{d\Omega_\nu}
    {\gamma^2(1-\beta\cos\theta_\nu)^2}.
    \label{eq:app_solid_angle_jacobian}
\end{align}

For an unpolarized muon decay, the neutrino angular distribution in the muon
rest frame is isotropic.
Using \cref{eq:app_solid_angle_jacobian}, the lab-frame angular distribution from a muon traveling horizontally at the origin $(\theta_\mu=0,z=0)$ is
\begin{align}
    \label{eq:app_boosted_angular_distribution_exact}
    \frac{dP_\nu}{d\Omega_\nu}(\theta_\mu=0,z=0)
    &=
    \frac{dP_\nu}{d\Omega^*_\nu}
    \frac{d\Omega^*_\nu}{d\Omega_\nu}
    \\\nonumber
    &=
    \frac{1}{4\pi}
    \frac{1}
    {\gamma^2(1-\beta\cos\theta_\nu)^2},
\end{align}
where the $\frac{dP_\nu}{d\Omega_\nu}$ is normalized to unity.

The result in \cref{eq:app_boosted_angular_distribution_exact} describes the idealized angular distribution in which the parent muon travels exactly along the nominal beam axis and decays at the origin. In a realistic beam, however, the neutrino distribution is centered around the local direction of the parent muon rather than the nominal beam axis. The observed forward-neutrino profile must therefore be obtained by convolving the boosted decay distribution with the phase-space distribution of the parent muon beam.

\begin{figure}[h]
    \centering
    \includegraphics[width=\linewidth]{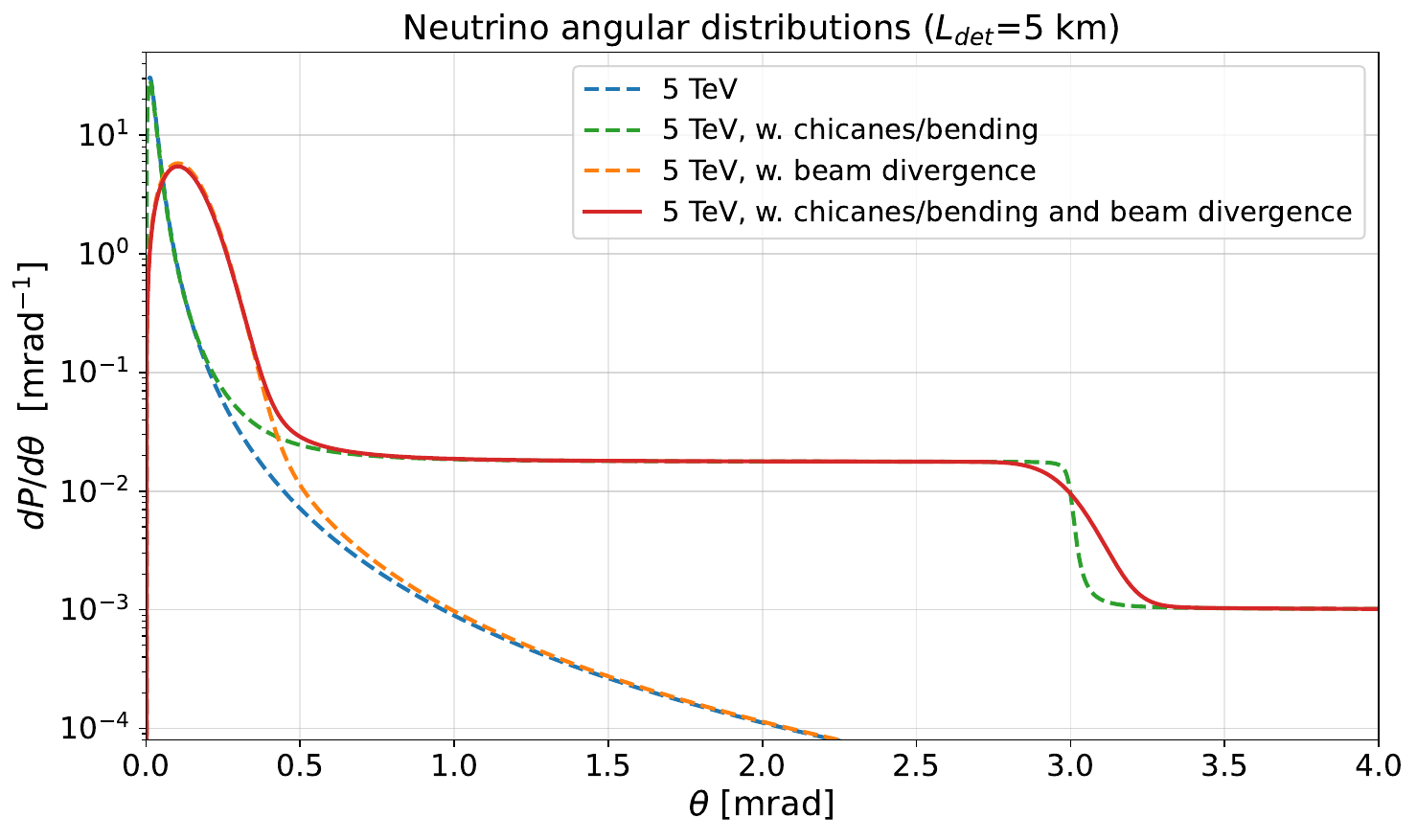}
    \caption{Angular distribution of the neutrino flux at a radius of 5 km away from IP. \textbf{Blue:} a 360 meter straight section with no chicanes/bending and no beam dynamics. \textbf{Green:} a 360 meter straight section including chicanes/bending but no beam dynamics. \textbf{Orange:} a 360-meter straight section with no chicanes/bending but including beam divergence assumed by Gaussian transverse angular distribution, $\theta_x,\theta_y\sim {N}(0,\sigma=0.1)$. \textbf{Red:} a 360 meter straight section with both chicanes/bending and beam divergence.
    The idealized boosted-decay result gives a narrow forward cone, while the inclusion of beam divergence broadens the distribution observed by a downstream detector. The plateau-like feature at the mrad level originates from muon decays inside the chicane, where the local beam direction varies along the lattice.
    \label{fig:angular_distribution_bd}}
\end{figure}

The beam divergence near the interaction point, ${\theta}_\mu \simeq {p}_{\mu,\perp}/p_{\mu,z}$, has the dominant effect on the neutrino profile as shown by \cref{fig:angular_distribution_bd}. 
For a multi-TeV muon beam, the intrinsic neutrino cone has a characteristic angular size $1/\gamma\sim 10^{-5}$~rad, while the angular spread of the parent muons can be substantially larger. 
In this regime, the neutrino direction relative to the nominal beam axis is primarily controlled by the local muon direction so the beam divergence takes over and broadens the neutrino beam profile.

The geometry of the beamline also plays a major role in shaping the neutrino beam. 
In chicanes and bending sections, the local tangent of the central orbit deviates from the nominal beam axis, allowing neutrinos to populate larger downstream angles and radii. 
In \cref{fig:angular_distribution_bd} we see this reflected in the plateau-like features in the angular distribution. Neutrinos at higher radii/transverse angles originate from the arcs and those in the middle (below 3 mrad) come from the chicanes.

The transverse size of the parent muon beam also smears the neutrino position at the observation surface. However, the beam size around the interaction region is typically at the $\mathcal{O}(\mathrm{cm})$ level, which is much smaller than the transverse scale associated with the angular smearing considered here. 
Similarly, the spread of decay positions along the straight section has only a subleading effect, since the neutrinos emitted at different $z$ positions propagate approximately along the same downstream direction with a 5-TeV boost.

For the angular distribution observed on a surface located a distance $L_{\rm det}$ from
the interaction point, the effect of beam divergence and decay-position spread
may be viewed as
\begin{align}
    &\left.
    \overline{\frac{dP_\nu}{d\Omega_{\nu}}}
    \right|_{L_{\rm det}} \nonumber \\
    \simeq
    &\int_{-L/2}^{L/2} dz\, f_{\rm dec}(z)
    \int d{\theta}_\mu\,
    f_\mu({\theta}_\mu;z)
    \left.\frac{dP_\nu}{d\Omega_{\nu}}(\theta_\mu,z)\right|_{L_{\rm det}}\,.
    \label{eq:angular_smearing_schematic}
\end{align}
Here, $dP_\nu/d\Omega_{\nu}$ denotes the lab-frame neutrino angular distribution
projected onto an observation sphere of radius $L_{\rm det}$ centered at the interaction
point, for a parent muon decaying at position $z$ and moving along the local
direction $\theta_\mu$. $f_\mu({\theta}_\mu;z)$ is the local angular distribution of the muon beam at position $z\in[-L/2,L/2]$ where $L$ is the total length of our interaction region (shown in the bottom panel of \cref{fig:beam_envelope}). 
The longitudinal distribution of decay positions follows from the exponential
decay of the muon population as the bunch travels through the straight section
\begin{align}
    f_{\rm dec}(z)
    =
    \frac{
    \exp\left[-(z+L/2)/\lambda_\mu\right]
    }{
    \lambda_\mu
    \left(1-\exp[-L/\lambda_\mu]\right)
    },
    \quad
    \lambda_\mu=\gamma c\tau_\mu ,
    \label{eq:decay_position_distribution}
\end{align}
where it is normalized over $z\in[-L/2,L/2]$.

\begin{figure*}[t]
    \centering
    \includegraphics[width=0.49\textwidth]{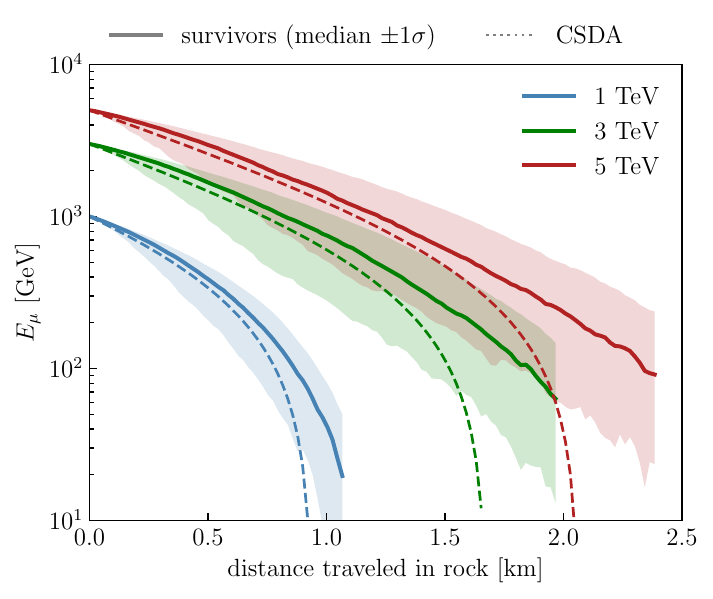}
    \includegraphics[width=0.49\textwidth]{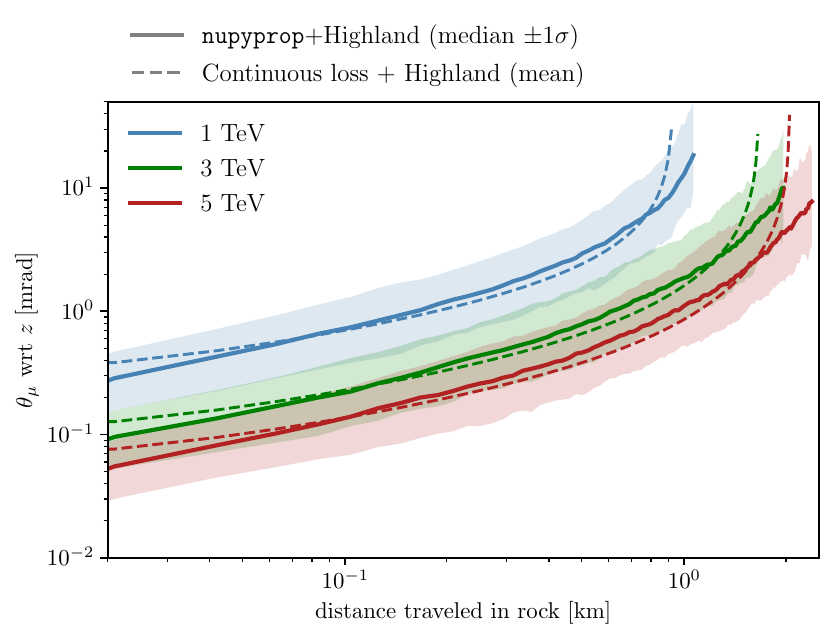}
    \caption{The energy loss (\textbf{left}) and angular deflection (\textbf{right}) of muons propagating through standard rock as a function of their travel distance.
    We inject muons at $1$~TeV, $3$~TeV, and $5$~TeV and propagate them through rock using \texttt{nuPyProp}~\cite{Garg:2022ugd}.
    The solid line corresponds to the median behavior of the muons that survive up to the given distance in our simulations ($E_\mu^{\rm sim} > 10$~GeV) and the dashed line is the mean of all muons and the continuous slowing down approximation (CSDA) mean expectation.
    The band corresponds to the $1\sigma$ spread in our simulations.
    \label{fig:muon_energy_and_angle}
    }
\end{figure*}


\section{Beam Dynamics }
\label{app:beam_optics}

The motion of the muons in the transverse plane is described by the Courant-Snyder (CS) formalism~\cite{Courant:1958wbj} (see \cite{Wiedemann:2007zbj} for a review).
The transverse coordinates of a muon are $u(s)$ and $u'(s)$, where $s$ is the position along the central reference orbit, $u=x,y$ are the transverse position, and $u'=du/ds = \tan \theta$ their rate of change with $s$. 
The angle $\theta$ is the angle of a given muon trajectory with respect to the reference orbit's tangent.
The beam is only ever deflected by small angles, so the paraxial approximation applies $u' = \tan \theta \simeq \theta \ll 1$.

The motion of the beam in the transverse plane obeys Hill's equation,
\begin{equation}
    u'' + K_u(s) \, u = 0, \qquad u=x,y ,
\end{equation}
where $K_u(s)$ is called the focusing strength.
If $K_u(s) = K_u$ is a constant, the solution is a simple harmonic oscillator, corresponding to the usual betatron oscillations.
The solution to the general case can be written as
\begin{equation}
    u(s) = \sqrt{\varepsilon_u \beta_u(s)} \cos\left(\psi_u(s)+\phi_u\right) ,
\end{equation}
where $\psi_u(s)$ is called the phase advance and $\phi_u$ is a constant phase.
The geometric emittance $\varepsilon_u$ is approximately a constant of motion throughout the ring and assumed to be the same in both transverse planes, $\varepsilon_x = \varepsilon_y = \varepsilon$.
We take the normalized transverse emittance at the 10 TeV MuC to be $\varepsilon_N = 25~\mu{\rm m\cdot rad}$~\cite{InternationalMuonCollider:2024jyv}, so that $\varepsilon = \varepsilon_N/\gamma_\mu$ for a relativistic muon beam.

The local Gaussian widths in each transverse plane are given by the envelopes of the CS solutions,
\begin{equation}
\label{eq:cs_envelopes}
    \sigma_u^2 = \varepsilon_u \beta_u, \quad \sigma_{u'}^2 = \varepsilon_u \gamma_u, \quad \langle u u' \rangle = -\varepsilon_u \alpha_u ,
\end{equation}
where
\begin{equation}
    \alpha_u = -\frac{1}{2}\beta_u',
    \quad
    \gamma_u = \frac{1+\alpha_u^2}{\beta_u}.
\end{equation}
The parameters $\alpha_u$, $\beta_u$, and $\gamma_u$ are called the CS or Twiss parameters.

The angular spread of the parent muons is determined locally by $\sigma_{u'}$, and is especially large at the interaction point, where the beam must be squeezed to the smallest size in order to maximize the collider luminosity.
For the latest designs, the beta function at the interaction point of the 10 TeV MuC is $\beta^* = 1.5$~mm, corresponding to a $\sigma_{x,y}\simeq 1$~$\mu$m~\cite{InternationalMuonCollider:2024jyv}.
For a 5 TeV muon beam, this gives 
\begin{equation}
    \sigma_{u'}^* = \sqrt{\frac{\varepsilon_N}{\gamma_\mu\beta^*}} \simeq 0.6~{\rm mrad},
\end{equation}
which is much larger than the characteristic decay angle $1/\gamma_\mu = m_\mu / E_\mu \simeq 0.02~{\rm mrad}$ (see, for example, \cref{fig:beam_envelope}).
The straight section, however, is much longer than the drift section between the final focusing magnets.
In the current design, the angular divergence for the longest drift section, $\sim 150$~m, is about $0.1 - 0.2$~mrad, which is the more appropriate scale for the angular divergence of the neutrino beam.

We do not track the muons individually through the lattice in \mint, but instead take the design optics as given and impose the envelopes in \cref{eq:cs_envelopes} at each point along the central orbit determined by the dipole magnet kicks. 
Quadrupole magnets focus in one transverse plane and defocus in the other according to the tabulated functions $\beta_{x,y}$ and $\alpha_{x,y}$.
Sextupoles are used to correct chromatic effects, but we do not include them in our simulation.
Dispersion is included in the horizontal plane, with the local offsets and angles given by $D_x(s)$ and $D'_x(s)$.

The MAD-X table is given as a discrete set of elements, while muon decays occur continuously along the ring.
Interpolating between the lattice elements is an important step, as a piece-wise interpolation of $\beta(s)$ is not adequate near low-$\beta$ regions, where the beam size changes very rapidly.
In drift regions, where the beam is allowed to propagate freely for a distance $d$, the linear optics has an exact solution:
\begin{equation}
    \beta_u(d) = \beta_{u}(0) - 2\alpha_{u}(0)d + \gamma_{u}(0) d^2,
\end{equation}
with $\gamma_u$, and consequently the beam divergence $\sigma_{u'}$, being constant over the drift. 
We reconstruct the beam size using this parabola, and apply the CS functions to find the angular divergence.
This accounts for the tabulated points and reproduces the expected $\beta(s)  = \beta^*+s^2/\beta^*$ behavior in the vicinity of the IP.
When discretizing the geometry of the central orbit, we subdivide the angular kicks evenly so that the total bending angle matches the value in the MAD-X table.

The longitudinal distribution of the muon bunch is not determined by the procedure above.
In our simulations, we assume a Gaussian momentum spread
\begin{equation}
\sigma_\delta \equiv \sigma_p/p_0 = 10^{-3},
\end{equation}
and the bunch length
\begin{equation}
\sigma_z = 1.5~{\rm mm},
\end{equation}
corresponding to a time spread of about $5$~ps. 
Off-momentum particles follow displaced orbits,
\begin{equation}
u_\delta = D_u \delta, \qquad u'_\delta = D'_u \delta.
\end{equation}
The horizontal beam widths are therefore modified as
\begin{equation}
\sigma_x^2 \to \varepsilon_x \beta_x + (D_x\sigma_\delta)^2,
\qquad
\sigma_{x'}^2 \to \varepsilon_x \gamma_x + (D'_x\sigma_\delta)^2 .
\end{equation}
The interaction region is designed to be dispersion-free, so this correction does not affect the most forward flux from the straight section.
In the matching sections, however, the dispersive contribution can dominate the
local angular spread.

We use as input a standard MAD-X~\cite{grote2003madx} output in TFS format.
Each row contains a lattice element, listing the position $s$ along the central reference orbit, the element length, the dipole bending angle, and the focusing properties of the lattice. 
For each decay, \mint samples the longitudinal position along the ring with the appropriate muon survival weight. 
The muon momentum is sampled from a Gaussian distribution centered at $p_0$ with width $\sigma_\delta p_0$. 
The transverse offsets and angles are then sampled from Gaussian distributions with the local widths determined by the CS functions and dispersion. 
In the present implementation we neglect the transverse correlations which go as $\langle u u'\rangle = -\varepsilon_u \alpha_u$ since they become negligible at sufficiently large distances.

\begin{figure}[t]
    \centering
    \includegraphics[width=\linewidth]{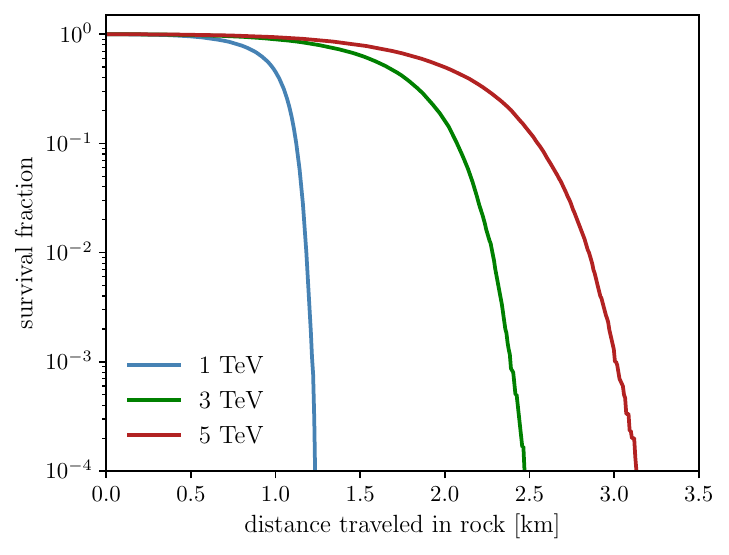}
    \caption{The survival probability of muons propagating through standard rock as a function of their travel distance.
    We inject muons at $1$~TeV, $3$~TeV, and $5$~TeV and propagate them through rock using \texttt{nuPyProp}~\cite{Garg:2022ugd}. 
    The survival probability is defined as the fraction of muons that reach a given distance with energy above $E_\mu^{\rm min}=10~\mathrm{GeV}$.
    \label{fig:muon_survival_probability}
    }
\end{figure}

\section{Muon Energy Loss in Rock}
\label{app:muon_energy_loss}

In this appendix, we show some additional validation plots for our muon energy loss and multiple scattering of muons in rock using \texttt{nuPyProp}~\cite{Garg:2022ugd}.
As a validation of our method, we inject monochromatic muons into a uniform slab of rock and track their energy, angle, and survival probability.

\Cref{fig:muon_energy_and_angle} shows the energy and angle as a function of their travel distance in rock for $10^3$ muons with energies of $1$~TeV, $3$~TeV, and $5$~TeV.
The left panel shows the median energy of the muons as a function of distance, while the right panel shows the median and mean of the angle of the muons as a function of distance.
The spread of the individual simulations is shown as a shaded band around the average.
\Cref{fig:muon_survival_probability} then shows the probability of the $1$~TeV, $3$~TeV, and $5$~TeV muons surviving with more than $E_\mu^{\rm min}=10$ GeV with respect to their distance traveled in standard rock.

\begin{figure*}[t]
    \centering
    \includegraphics[width=0.49\textwidth]{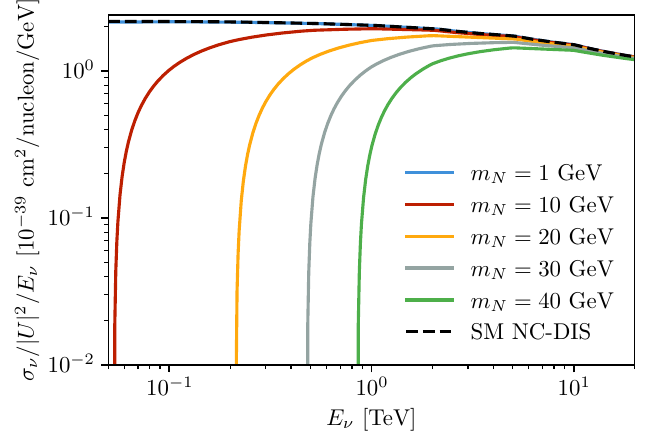}
    \includegraphics[width=0.49\textwidth]{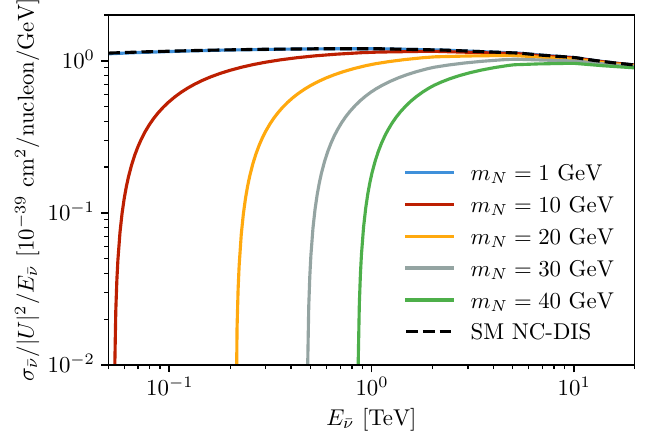}
    \caption{
       The upscattering neutrino (\textbf{left}) and antineutrino (\textbf{right}) cross-sections with respect to energy for different HNL masses (solid color lines), normalized by the mixing matrix element squared. 
       The dashed line shows the SM NC DIS case.
    \label{fig:upscattering_xsec}}
\end{figure*}

Between stochastic energy loss events, the energy loss is evolved according to the continuous slowing down approximation (CSDA)~\cite{Lipari:1991ut}:
\begin{equation}
\label{eq:secondary_muon_continuous_loss}
\frac{dE_\mu}{dX}=-\alpha(E_\mu)-\beta_{\rm cut}(E_\mu)E_\mu,\qquad X=\rho z,
\end{equation}
with standard-rock density $\rho=2.65~\mathrm{g/cm^3}$. 
Catastrophic bremsstrahlung, pair production, and photonuclear losses with fractional energy loss $y>10^{-3}$ are sampled from the \texttt{nuPyProp} integrated-cross-section CDFs, using the ALLM photonuclear model.
Muon decay in flight is also included, but is typically negligible for TeV-scale muons with our stopping energy threshold.

To estimate the angular deflection, which is not simulated by
\texttt{nuPyProp}, we supplement the energy-loss propagation with a
simplified multiple-Coulomb-scattering model. For each propagation step, the projected angular kicks are sampled from Gaussian distributions with width given by the Highland expression~\cite{Highland:1975pq},
\begin{equation}
\theta_0=
\frac{13.6~\mathrm{MeV}}{\beta_\mu p_\mu}
\sqrt{\frac{X}{X_0}}
\left[1+0.038\ln\left(\frac{X}{X_0}\right)\right],
\label{eq:secondary_muon_highland}
\end{equation}
where $X$ is the column depth traversed in the step and $X_0$ is the radiation length of the medium. Equivalently, the polar deflection is Rayleigh distributed with scale $\theta_0$, while its azimuth is sampled uniformly in $[0,2\pi)$. This treatment captures the characteristic angular broadening from multiple Coulomb scattering, which dominates the accumulated deflection for high-energy muons. 
Dedicated propagation studies~\cite{Gutjahr:2022quk} provide the cumulative angular deflection distribution for a $\sim$TeV muon traveling through water, which is peaked around $\sim1~$mrad.

\begin{figure}[t]
    \centering
    \includegraphics[width=0.49\textwidth]{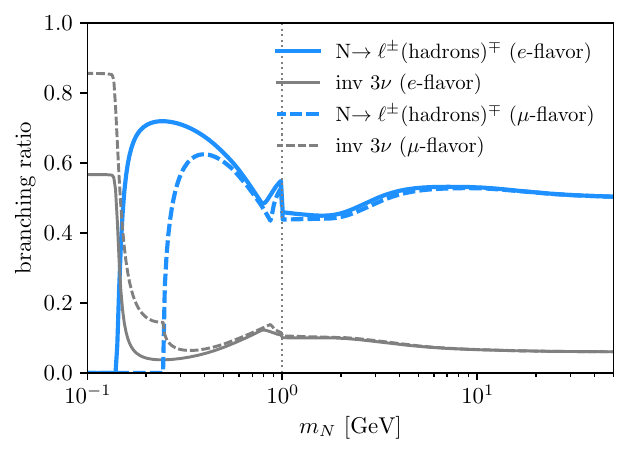}
    \caption{
       Decay branching ratios of HNLs mixed exclusively with $e$-flavor (\textbf{solid}) and $\mu$-flavor (\textbf{dashed}). The blue lines represent channels with visible final states and the gray lines represent the fully-invisible three-neutrino channel.
    \label{fig:hnl_branching}
    }
\end{figure}

The column density of our step sizes is $5 \times 10^{3}$~g/cm$^{2}$ ($\simeq 20$~m of rock), which is sufficient to resolve the muon stopping points. 
At few-GeV energies the stochastic interaction length grows beyond $10^{5}~\mathrm{g/cm^{2}}$ and an uncapped step would smear them by hundreds of meters. We validated the energy-loss treatment in two ways: the ensemble-mean energy (counting stopped muons at zero) reproduces the deterministic solution of \cref{eq:secondary_muon_continuous_loss} evaluated with the total $\beta$ to better than $2.5\%$ out to half the CSDA range for $1$ to $5$~TeV muons, and the survival probability at the CSDA range is $68\%$ ($55\%$) at $1$~TeV ($5$~TeV), the expected size of range straggling for radiative-loss-dominated propagation.

\section{HNL Cross-Sections}
\label{app:HNL_xsec}
\Cref{fig:upscattering_xsec} shows the cross-sections for neutrino (left) and antineutrino (right) upscattering on nucleons for isoscalar matter based on their energy. 
Each color represents the cross-section calculated using various masses for the HNL, where we can see that heavier HNLs have a larger energy threshold and that as $m_N$ decreases, the cross section tends to the SM NC cross section up to a mixing matrix element factor. 

\Cref{fig:hnl_branching} then shows the branching ratios of the neutrino decay modes we considered in \cref{sec:hnl}. 
We show the fully invisible $N\to \nu \nu \bar\nu$ channel as well as the fully visible $N \to \ell^\pm (\text{hadrons})^{\mp}$.
The remaining semi-visible channels, such as $N \to \ell^\pm \ell^\mp \bar\nu$, are not shown but also not included in the analysis of \cref{sec:hnl}.
We pick $m_N=1$~GeV as the transition from the low-energy picture to the high-energy one.
The former uses exclusive hadronic channels to calculate the branching ratios, while the latter calculates the hadronic branching ratios at the quark level inclusively.


\bibliographystyle{apsrev4-1}
\bibliography{references}

\end{document}